\documentclass[useAMS,usenatbib]{mnras}

\usepackage{graphicx} 
\usepackage{rotating}
\usepackage{subfig}

\def\kms{\mbox{km~s$^{-1}$}}
\def\HI{H\,{\sc i}}
\def\aap{A\&A}
\def\apj{ApJ}
\def\apjl{ApJL}

\def\mnras{MNRAS}

\def\aj{AJ}

\def\apjs{ApJS}

\title[The Distance and Properties of Hydrogen Clouds in the Leading Arm of the Magellanic System]
      {The Distance and Properties of Hydrogen Clouds in the Leading Arm of the Magellanic System}

\author[B.-Q. For, L. Staveley-Smith, N.~M. McClure-Griffiths, T. Westmeier, K. Bekki]
{B.-Q. For$^{1}$\thanks{E-mail:biqing.for@icrar.org}, 
L. Staveley-Smith$^{1}$, N.~M. McClure-Griffiths$^{2}$, 
\newauthor
T. Westmeier$^{1}$, K. Bekki$^{1}$\\
$^{1}$ICRAR, University of Western Australia, Crawley, WA, 6009, Australia\\
$^{2}$Research School of Astronomy \& Astrophysics, Australian National University, Canberra, ACT 2611, Australia}

\begin{document}

\date{Accepted 27 May 2016. Received 7 April 2016}

\pagerange{\pageref{firstpage}--\pageref{lastpage}} \pubyear{2015}

\maketitle

\label{firstpage}

\begin{abstract}
We present a high-resolution study of five high-velocity clouds in the Magellanic Leading Arm region. 
This is a follow-up study of our widefield Parkes survey of the region in order to 
probe the multiphase structures of the clouds and to give an insight to their origin, 
evolution and distance. 
High-resolution data were obtained from the Australia Telescope Compact Array. 
By combining with single-dish data from the 
Galactic All-Sky Survey (GASS), we are able to probe compact and diffuse emission simultaneously. 
We identify resolved and unresolved clumps. Physical 
parameters were derived for both diffuse structure and compact clumps. The latter are cold with 
typical velocity linewidths of 5~\kms. 
We find a gradient in thermal halo pressure, hydrogen density and \HI\ column density of HVC 
as a function of Galactic latitude. This is possibly the first observational evidence of varying distance 
in the Leading Arm region, 
with the leading part of the Leading Arm (LA II and III) probably being closer to the Galactic 
disc than the trailing end (LA I). 
\end{abstract}


\section{INTRODUCTION\label{intro}}

High-velocity clouds (HVCs; \citealp{Muller63}) are neutral atomic 
hydrogen gas clouds distributed across the entire sky as large, 
homogeneous complexes, and include the Magellanic Stream \citep{Mathewson74}, 
as well as numerous compact and isolated clouds. 
They are characterised by a high radial velocity that is 
forbidden by a simple Galactic rotation model. To classify HVCs, 
\citet{Wakker91} introduced the so-called ``deviation velocity'', which is 
defined as the smallest difference between the velocity 
of the cloud and that of the Galactic disc gas along the line-of-sight, 
and suggested a deviation 
velocity of at least 50~\kms\ for HVCs.

The origin of HVCs has been under debate for decades. 
A hypothesis of an extragalactic origin was put forward by \citet{Blitz99}. 
They argued that the observed properties of HVCs are consistent with a 
distribution across the entire Local Group. In fact, numerical simulations 
by \citet{Klypin99} and \citet{Moore99} 
suggested that HVCs, excluding some of the large HVC complexes 
and the Magellanic Stream (MS), might even be the missing dark matter satellites as 
predicted by cosmological dark matter models.    
\citet{BB99} conducted a study using the Leiden/Dwingeloo Survey of Galactic 
neutral hydrogen (LDS; \citealp{HB97}). They identified a subclass of HVCs, namely 
compact high-velocity clouds (CHVCs), which are isolated, have an angular diameter 
of $\leq2\degr$ and \HI\ column densities above $1.5\times10^{18}$~cm$^{-2}$. 
These CHVCs were again claimed to be at extragalactic distance. However, 
surveys of local galaxy groups 
showed no detection of similar \HI\ clouds \citep{Zwaan01,Pisano04}. 
A more detailed study of a subsample of CHVCs concluded that they are most likely 
circumgalactic objects with distances of the order of 100~kpc \citep{Westmeier05}. 

Many CHVCs have a head-tail morphology \citep{Bruns00,Putman11}, suggesting
that gas has been stripped via ram-pressure interaction with the ambient medium. 
A recent simulations of \citet{Salem15} has shown such ram-pressure effect on the Large Magellanic Cloud (LMC). 
Some of the head-tail clouds have been studied in detail, such as HVC~125+41-207 \citep{Bruns01}, 
HVC~289+33+251 \citep{Bruns04}, HVC~291+26+195 
and HVC~297+09+253 (\citealp{NBB06}, hereafter BBKW06). In many cases, they reveal 
two-component line profiles consisting of a cold core surrounded by a warm envelope.
Nevertheless, numerical studies are somewhat inconclusive about the existence of two-phase 
structures in HVCs with $z$ greater than $\sim20$~kpc in a $T\sim10^{6}$~K halo \citep{Wolfire95}. 
Observationally, two-phase HVCs have been found in the Magellanic Stream and Leading Arm (LA)  
(e.g. see \citealp{Stan08,For13,For14}), which have estimated distances between 20 and 100~kpc.  
This implies that the thermal pressure and halo density beyond 10~kpc is not well understood. 

In this paper, we study the multiphase structure of a subsample of HVCs in the region of the LA 
(\citealp{For13}, hereafter FSM13). 
The aims are (1) to resolve any multiphase structure, (2) derive physical parameters, 
(3) study physical properties, in particular the thermal pressure 
in different regions to probe environmental effects. 
The paper is organised as follows: in \S2, we list the selection criteria for the targets.  
\S3 gives a decription of our observational setup, data acquisition and reduction process. 
\S4 describes the analysis methodology. Our results and interpretation are presented in \S5. 
We compare the results with previous studies and discuss the implications for the Galactic halo 
by comparing our results with a model in \S6 and \S7. Finally, a summary is given in \S8.  

\section{TARGET SELECTION\label{selection}}

\begin{figure*}
\includegraphics[scale=0.6]{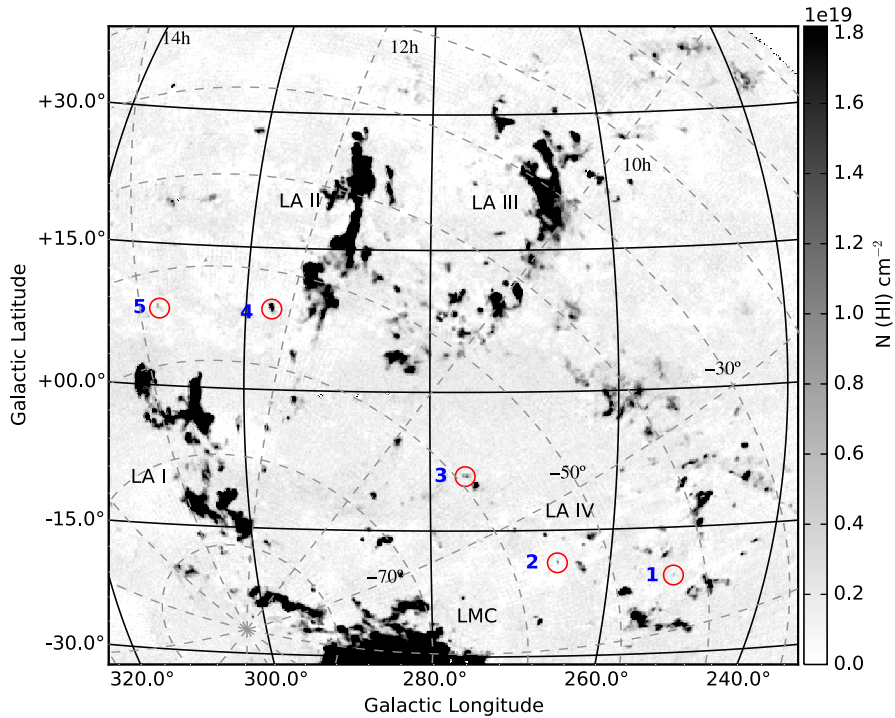}
\caption{Integrated \HI\ column density map in the region of the LA as studied by FSM13. 
The red circles mark the clouds selected for our ATCA observations at different Galactic latitudes. 
Equatorial coordinates are overlaid onto the plot in dashed grey. 
The labels 1--5 correspond to HVC~252.2-20.5+310, HVC~266.0-18.7+338, HVC~276.3-09.0+277, HVC~297.1+08.5+253 
and HVC~310.3+08.1+167, respectively. Three of the clouds are part of the LA IV.}
\label{position}
\end{figure*}

We selected five HVCs from the FSM13 catalog. The selection criteria are 
based on the following: (a) the peak brightness temperature of the HVC is $\sim10\sigma$ 
above the brightness 
temperature sensitivity for the instrument setup and array configuration (see \S3); 
(b) the HVC has a velocity linewidth of less than 10~\kms; (c) the HVC has a distinct 
morphological type of head-tail, bow-shock or symmetric as described in FSM13; 
(d) the HVCs are over a range of different Galactic latitudes. 
Figure~\ref{position} shows the integrated \HI\ column density map of the 
LA region studied by FSM13, and the observed targets are marked with red circles. 
 
\section{OBSERVATIONS AND DATA\label{obs}}

Both single-dish and interferometer data are used for this study. 
While single-dish data probe the diffuse emission, 
the interferometer data allow us to probe small-scale structures. 
We employed data from the second data release of the single-dish Parkes 
Galactic All-Sky Survey (GASS; \citealp{NMG09}) in this study. This 
has been corrected for stray radiation and radio frequency interference \citep{Kalberla10}. 
GASS is an HI survey of the entire sky south of declination +$1\degr$ 
using the 20-cm multibeam receiver on the Parkes radio telescope.
GASS covers $V_{\rm LSR}$ between $-$400 and +500~\kms.
The data have a channel width of 0.82~\kms, a spectral resolution of 1~\kms,
a brightness temperature ($T_{B}$) rms sensitivity of 57~mK, and an
angular resolution of 16$\arcmin$. We refer the reader to detailed descriptions 
of GASS and data processing in \citet{NMG09} and \citet{Kalberla10}. 

The high-resolution \HI\ observations were carried out 
at the Australia Telescope Compact Array (ATCA) in Narrabri 
using the EW367 and EW352 configurations in 
November 2012 and January 2013, respectively.   
Only five out of the six antennas were used for the analysis. 
The inner five antennas of the EW367 configuration cover baselines between 92~m and 367~m. 
The EW352 configuration covers baselines 
between 31~m and 352~m. We used the zoom mode option of the 
Compact Array Broadband Backend (CABB; \citealp{Wilson11}) 
for the observations, which has 
a total bandwidth of 2~GHz with 1~MHz (coarse) and 0.5~kHz (fine) resolutions. 
We centered the zoom band on 1420~MHz. 
This configuration results in a velocity resolution of 0.1~\kms.

We observed the primary flux calibrator PKS~1934-638 at the beginning of each 
observing session. A secondary phase calibrator, which is located nearby the 
target field, was observed for 3~min every hour. 
Each observing session was about 10--12 h long, so achieved good $uv$-coverage. 
We also adopted the strategy of obseving two targets per observing session 
by alternating between them. The beam was centered on the coordinates 
listed in the FSM13 catalog.  

The data reduction was performed in a standard manner 
using MIRIAD \citep{Sault95}. To create dirty images, we 
adopted robust weighting of 0 for four of the HVCs, 
which optimizes resolution 
and sensitivity. For HVC~310+08+167, a robustness parameter of 2 was used  
in order to match for its low surface brightness. 
A small selected region with \HI\ emission in each dirty cube was 
deconvolved using the Steer CLEAN algorithm \citep{Steer84}. The final data cubes were corrected 
for primary beam attenuation and have an rms noise of about 10~mJy per beam.

\begin{figure*}
\center
\includegraphics[scale=0.42]{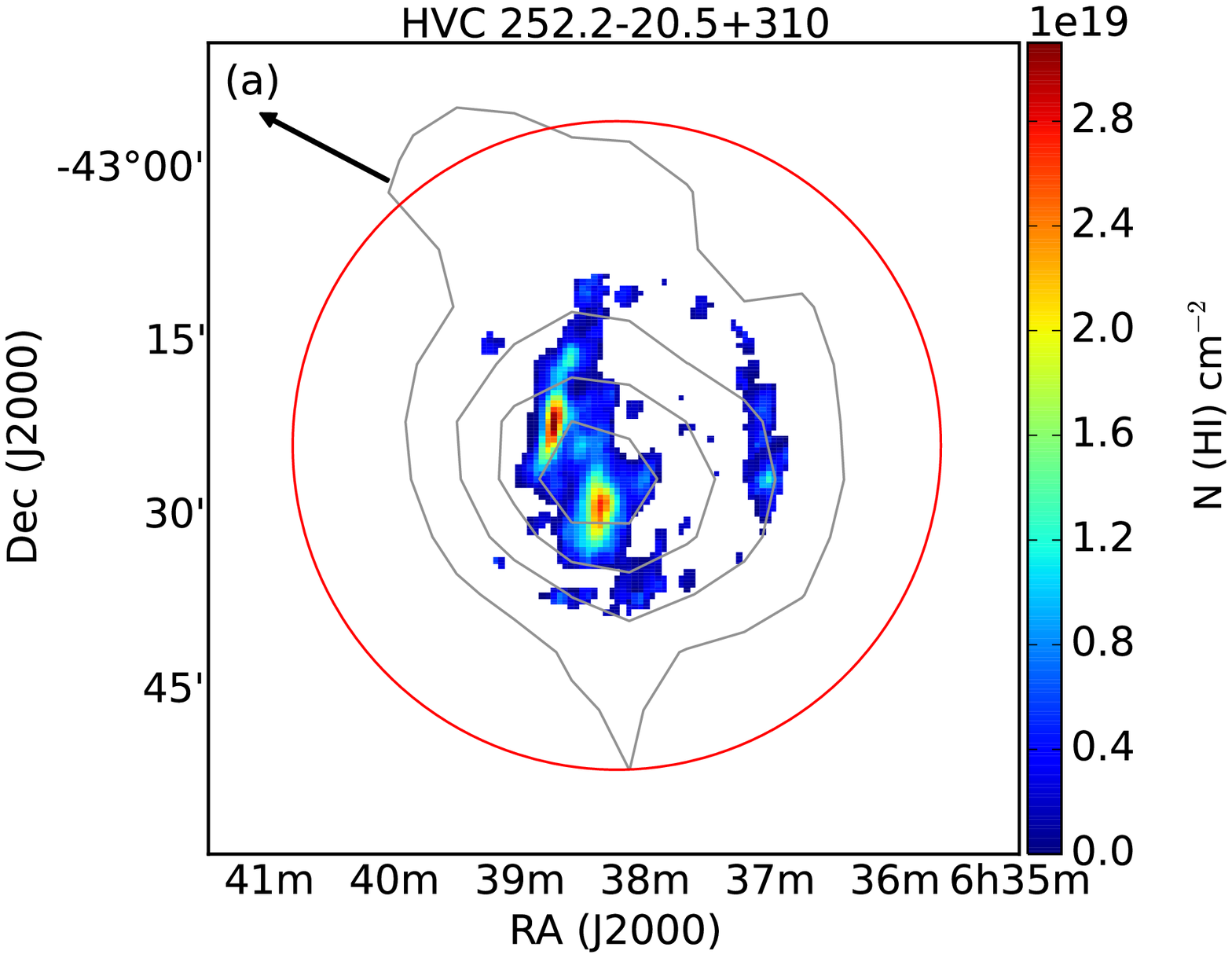}
\includegraphics[scale=0.42]{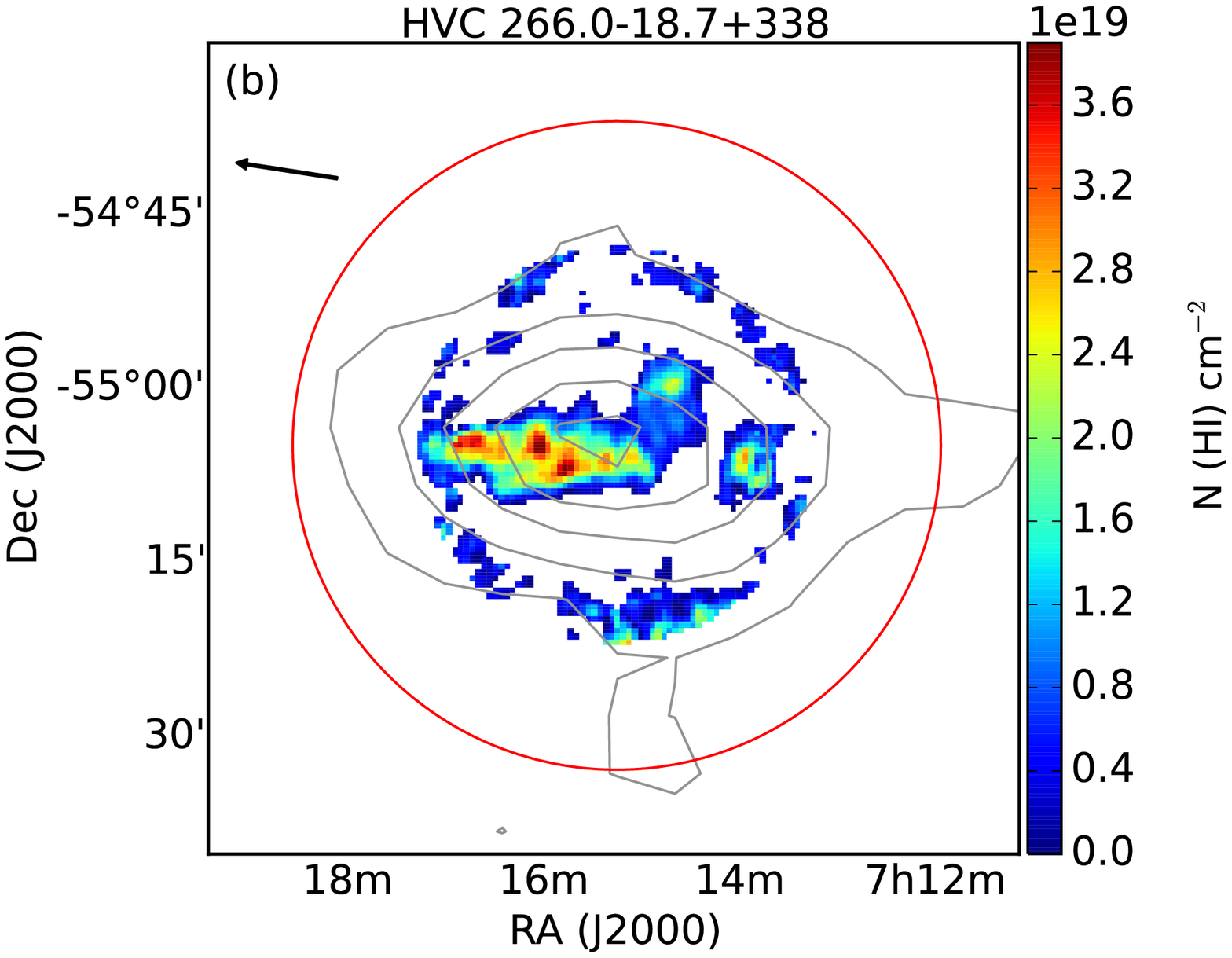}
\includegraphics[scale=0.42]{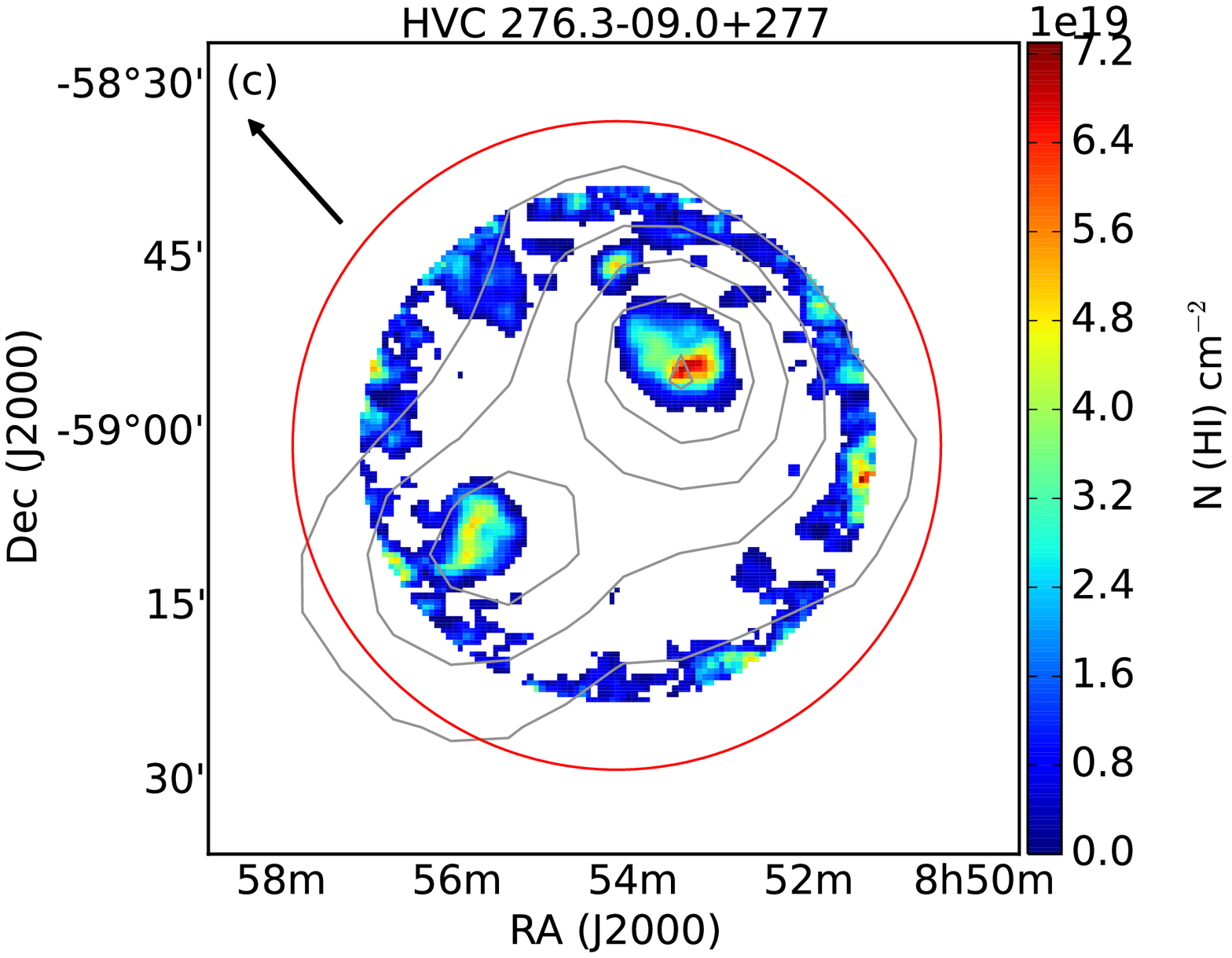}
\includegraphics[scale=0.42]{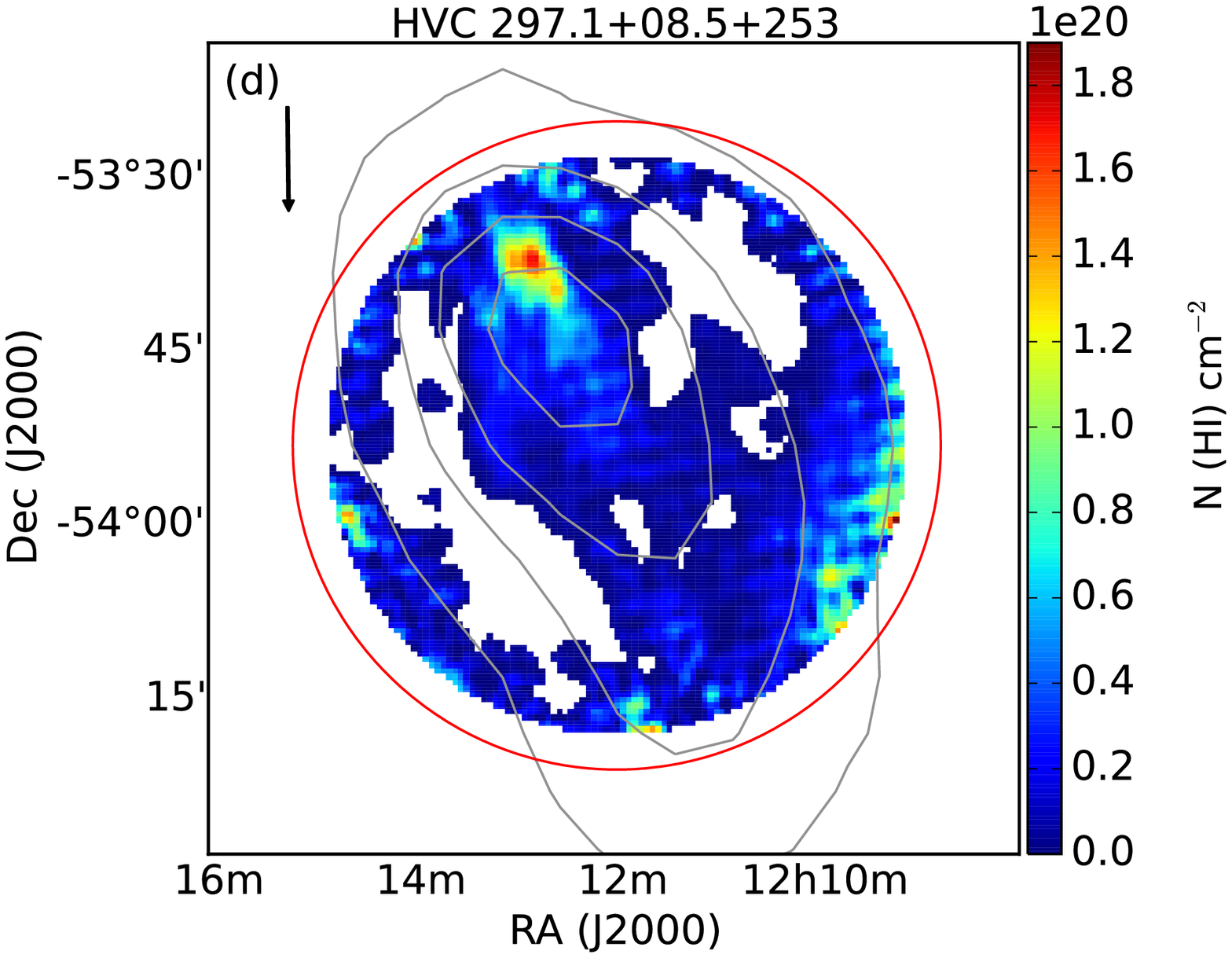}
\includegraphics[scale=0.42]{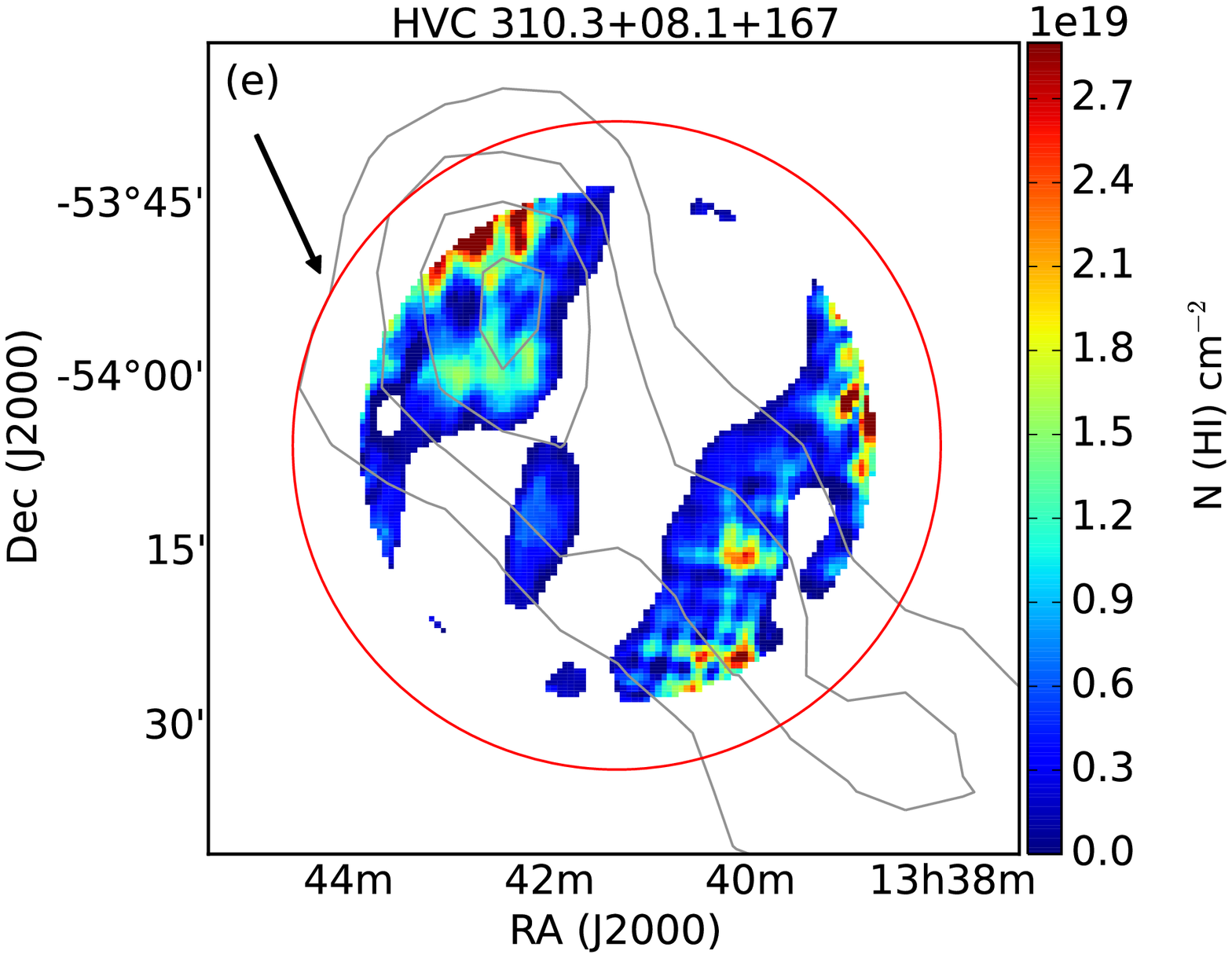}
\caption{Integrated \HI\ column density maps of the combined Parkes and 
ATCA \HI\ data set. The red circles indicate the 12 percent of peak sensitivity area of the image. 
The analyzed regions as shown here are based on the percentage of the peak sensitivity. 
The grey contours represent \HI\ column densities in the GASS data set. The arrows point in the direction of the Galactic plane.}
\label{merged_GASScon}
\end{figure*}

To circumvent the short-spacing problem of the interferometer data, we combined 
the ATCA data with the single-dish data from GASS. 
We adopted the linear method, merging in the Fourier domain \citep{Stan02}. 
With this method, the GASS data were converted to Jy per beam 
with a conversion factor of 0.658, 
regridded spatially and in velocity to match the ATCA image, and then 
the residual primary beam attenuation present in the 
ATCA data was applied to the GASS data prior to combining them. 
We smoothed the combined data cube spatially using the CONVOL task. The cubes 
are smoothed to full width half maxima (FWHM) of 
150\arcsec, 150\arcsec, 160\arcsec, 200\arcsec\ and 300\arcsec\ for HVC~252.2$-$20.5+310, 
HVC~266.0$-$18.7+338-18, HVC~276.3$-$09.0+277, HVC~297.1+08.5+253, HVC~310.3+08.1+167, respectively. 
We also smoothed the cubes spectrally over 3 channels using a Hanning function and 
extracted the channels that contain \HI\ emission.   
As the sensitivity drops off quickly toward the outer edge of the beam, 
we used the ATCA primary beam model to mask out the noise at the outer edge 
of the data cube. 
Then, we created the 
integrated \HI\ column density maps of 5 clouds 
as shown in Figure~\ref{merged_GASScon}. 

\section{Analysis\label{analysis}}

\begin{figure*}
\center
\includegraphics[scale=1.0]{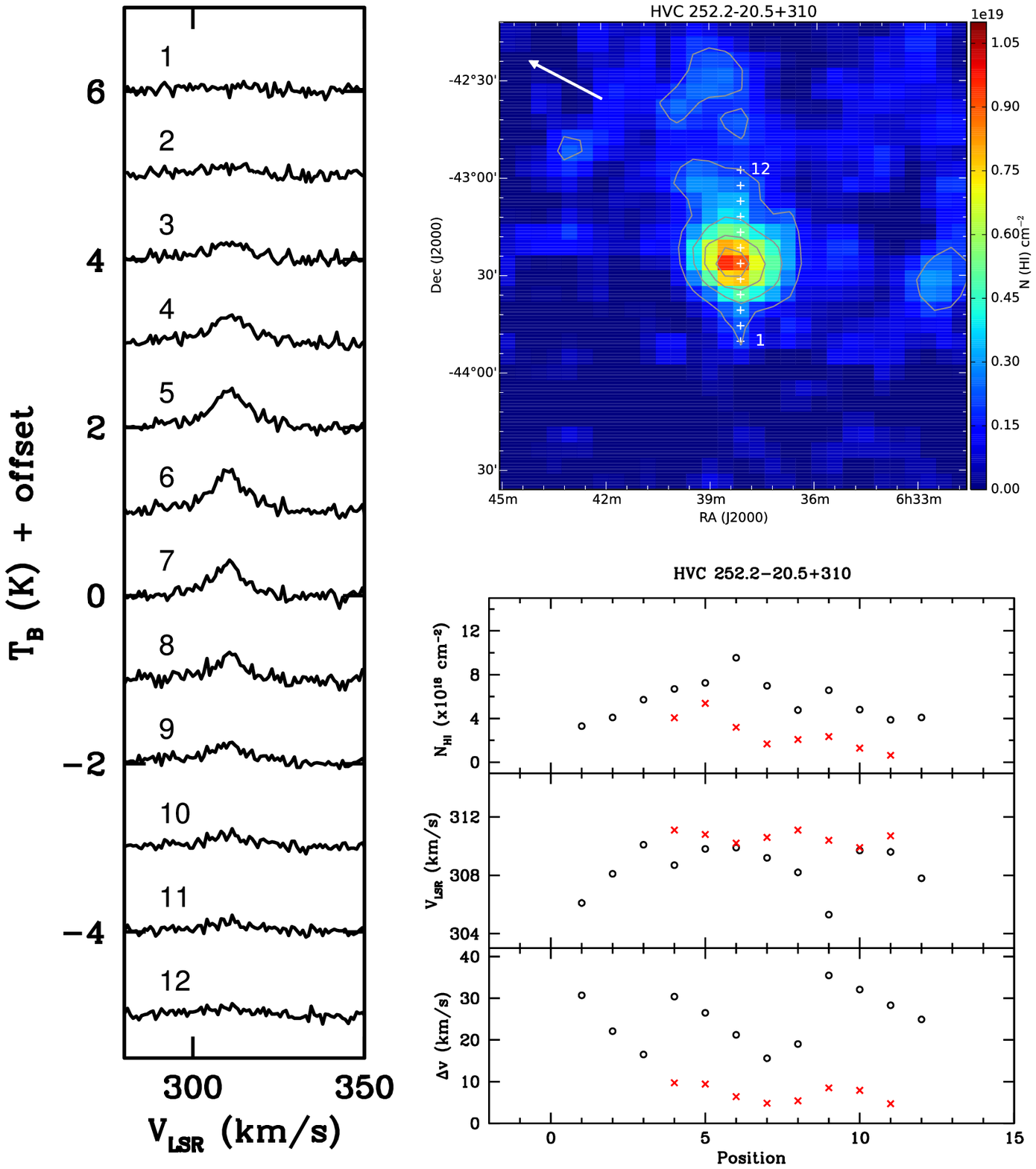}
\caption{{\bf Top right:} Integrated \HI\ column density map of HVC~252.2-20.5+310 in the GASS data set. 
The contour levels correspond to \HI\ column densities ($N_{\rm HI}$) of 0.2, 0.4, 0.6, 0.8, 1.0$\times$10$^{19}$~cm$^{-2}$. 
The white crosses represent the positions along the sliced axis of the cloud. 
It is sliced from south to north with a designated position number. The arrow 
points in the direction of the Galactic plane.
{\bf Left:} Series of extracted spectra along the sliced axis. 
{\bf Bottom right:} Physical parameters of the cloud as derived from the spectra. 
Black circles and red crosses represent 
the broad and narrow component, respectively.}
\label{Gcomb1}
\end{figure*}

\begin{figure*}
\center
\includegraphics[scale=1.0]{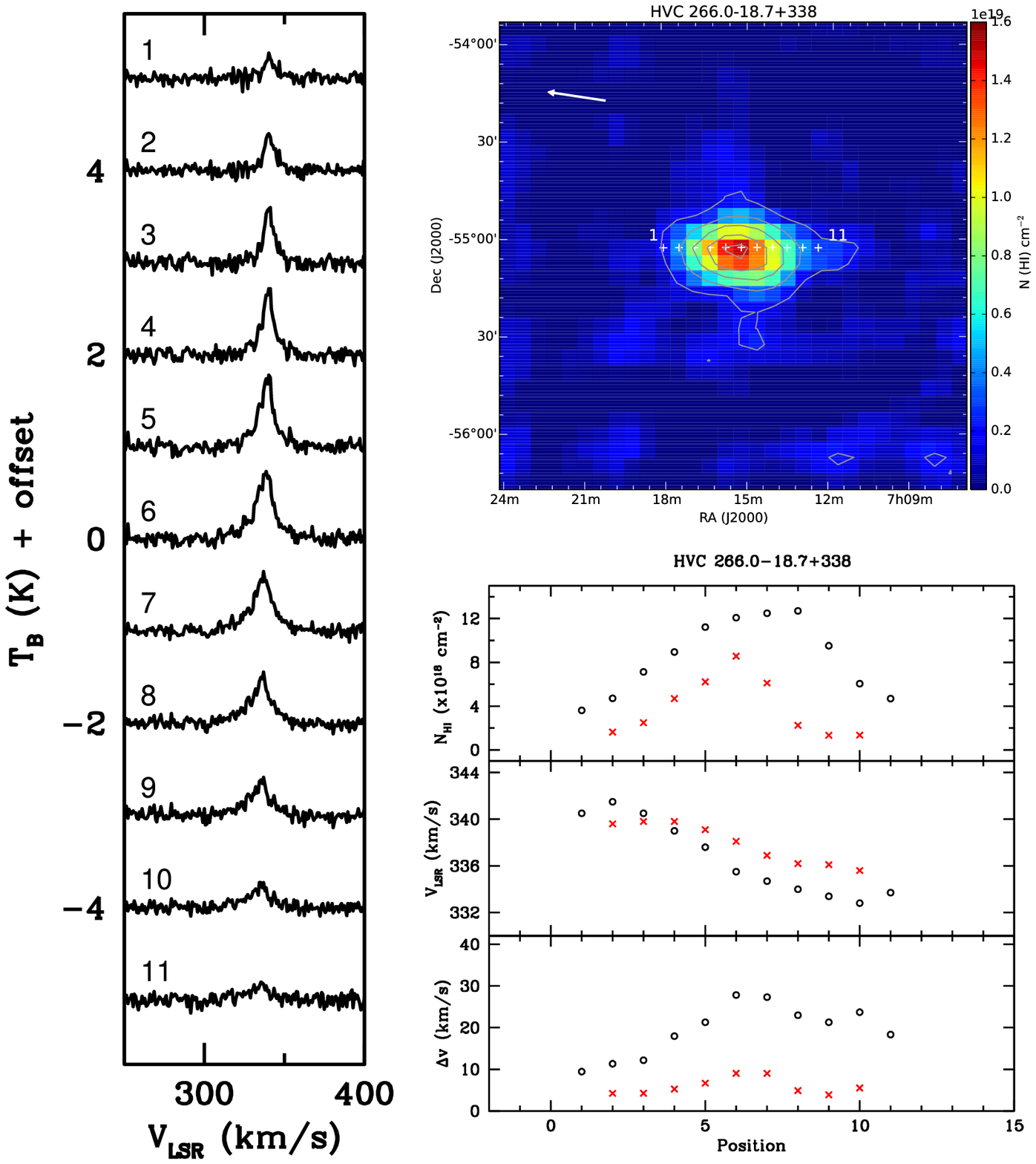}
\caption{Same as Figure~\ref{Gcomb1} except for HVC~266.0-18.7+338. 
The contour levels correspond to 0.2, 0.5 , 0.8, 1.1, 1.40$\times$10$^{19}$~cm$^{-2}$ 
and the cloud is sliced from east to west.}
\label{Gcomb2}
\end{figure*}

\begin{figure*}
\center
\includegraphics[scale=1.0]{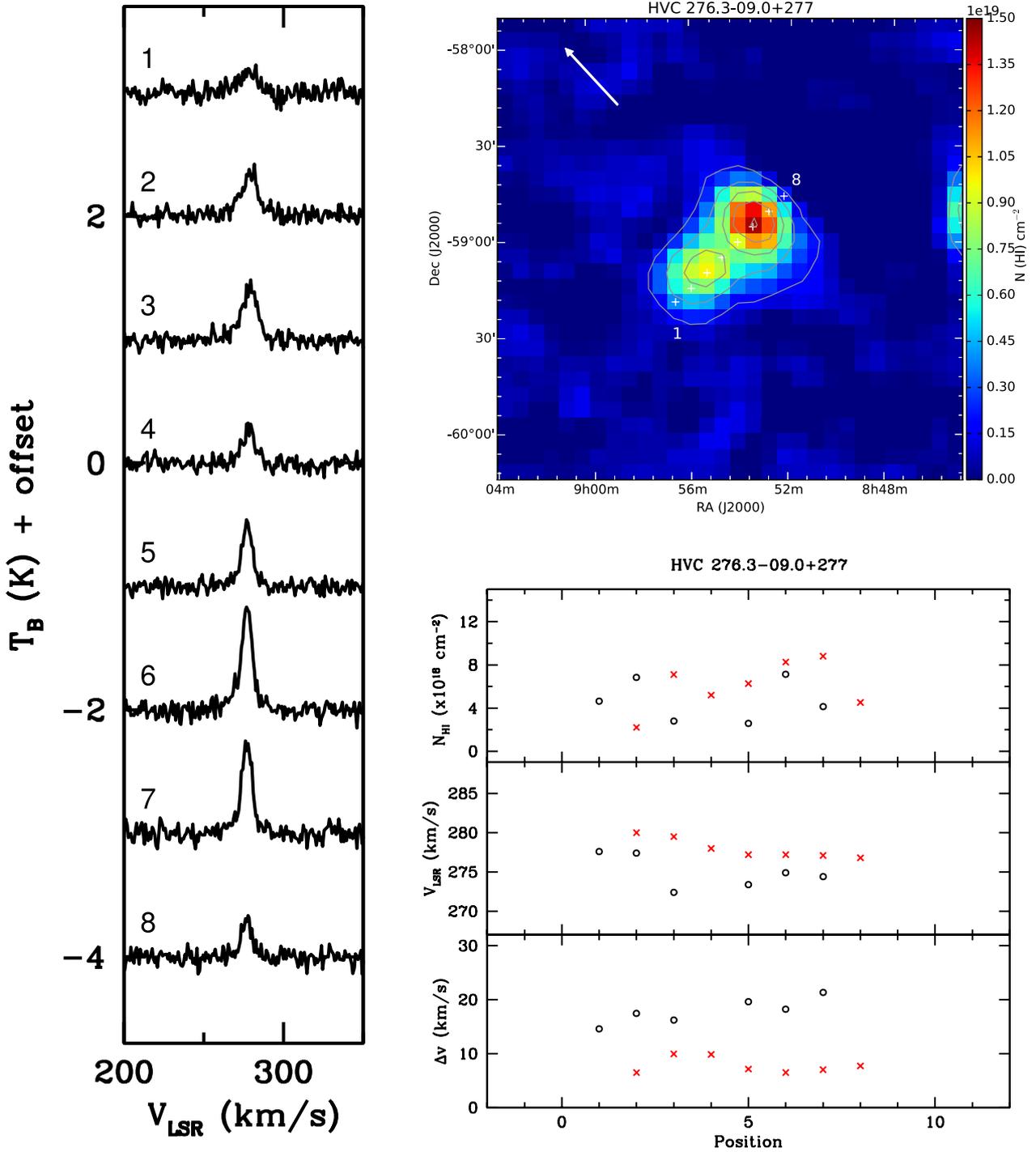}
\caption{Same as Figure~\ref{Gcomb1} except for HVC~276.3-09.0+277. 
The contour levels correspond to 0.2, 0.5 , 0.8, 1.1, 1.40$\times$10$^{19}$~cm$^{-2}$ 
and the cloud is sliced from south-east to north-west.}
\label{Gcomb3}
\end{figure*}

\begin{figure*}
\center
\includegraphics[scale=1.0]{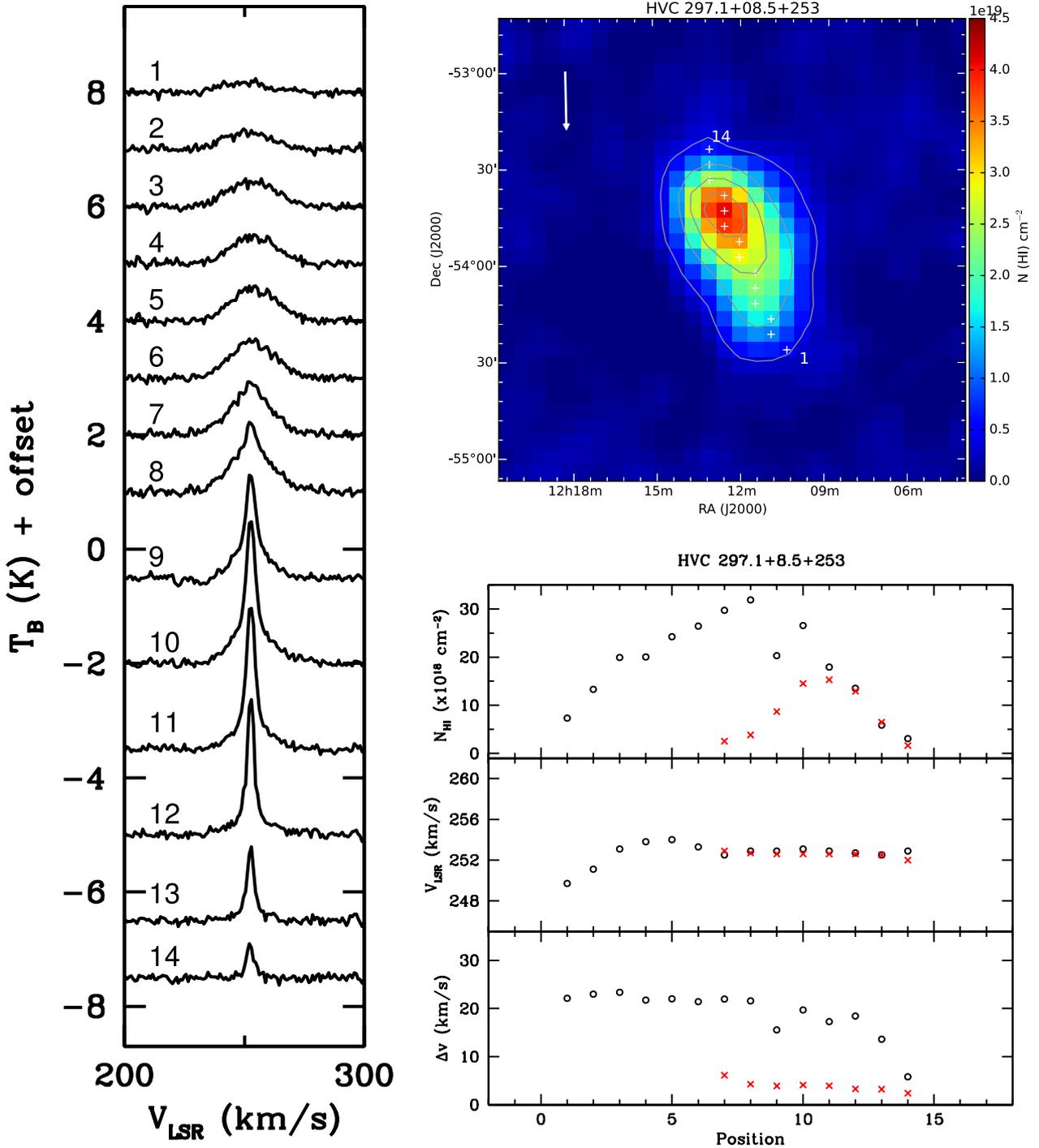}
\caption{Same as Figure~\ref{Gcomb1} except for HVC~297.1+08.5+253. 
The contour levels correspond to 0.5, 1.5, 2.5, 3.5, 4.5$\times$10$^{19}$~cm$^{-2}$ and the cloud is sliced from south-west to north-east. }
\label{Gcomb4}
\end{figure*}

\begin{figure*}
\center
\includegraphics[scale=1.0]{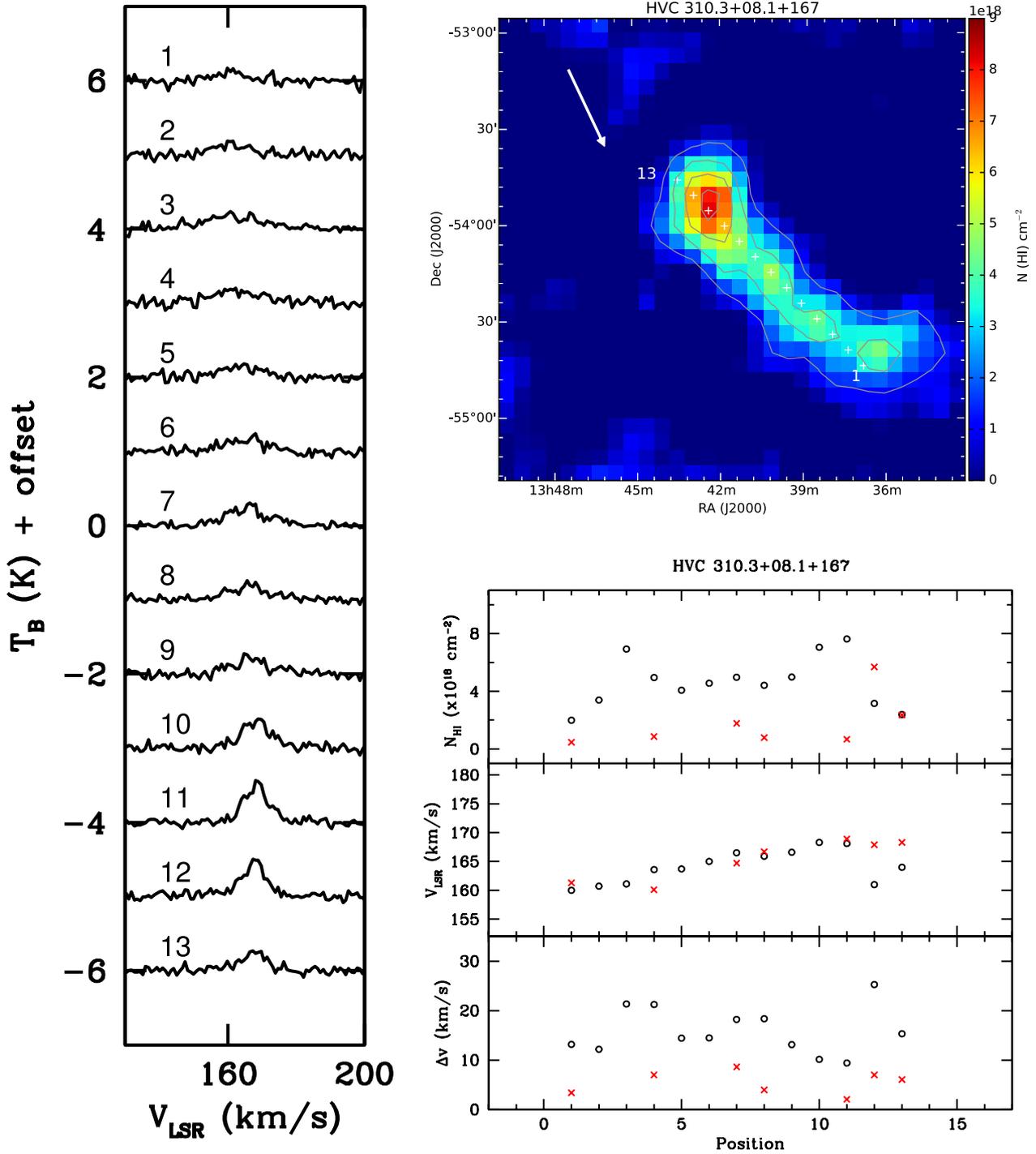}
\caption{Same as Figure~\ref{Gcomb1} except for HVC~310.3+08.1+167. 
The contour levels correspond to 1.5, 3.5, 5.5, 7.5, 9.5$\times$10$^{18}$~cm$^{-2}$ and 
the cloud is sliced from south-west to north-east.}
\label{Gcomb5}
\end{figure*}

\begin{figure*}
\center
\includegraphics[scale=0.4]{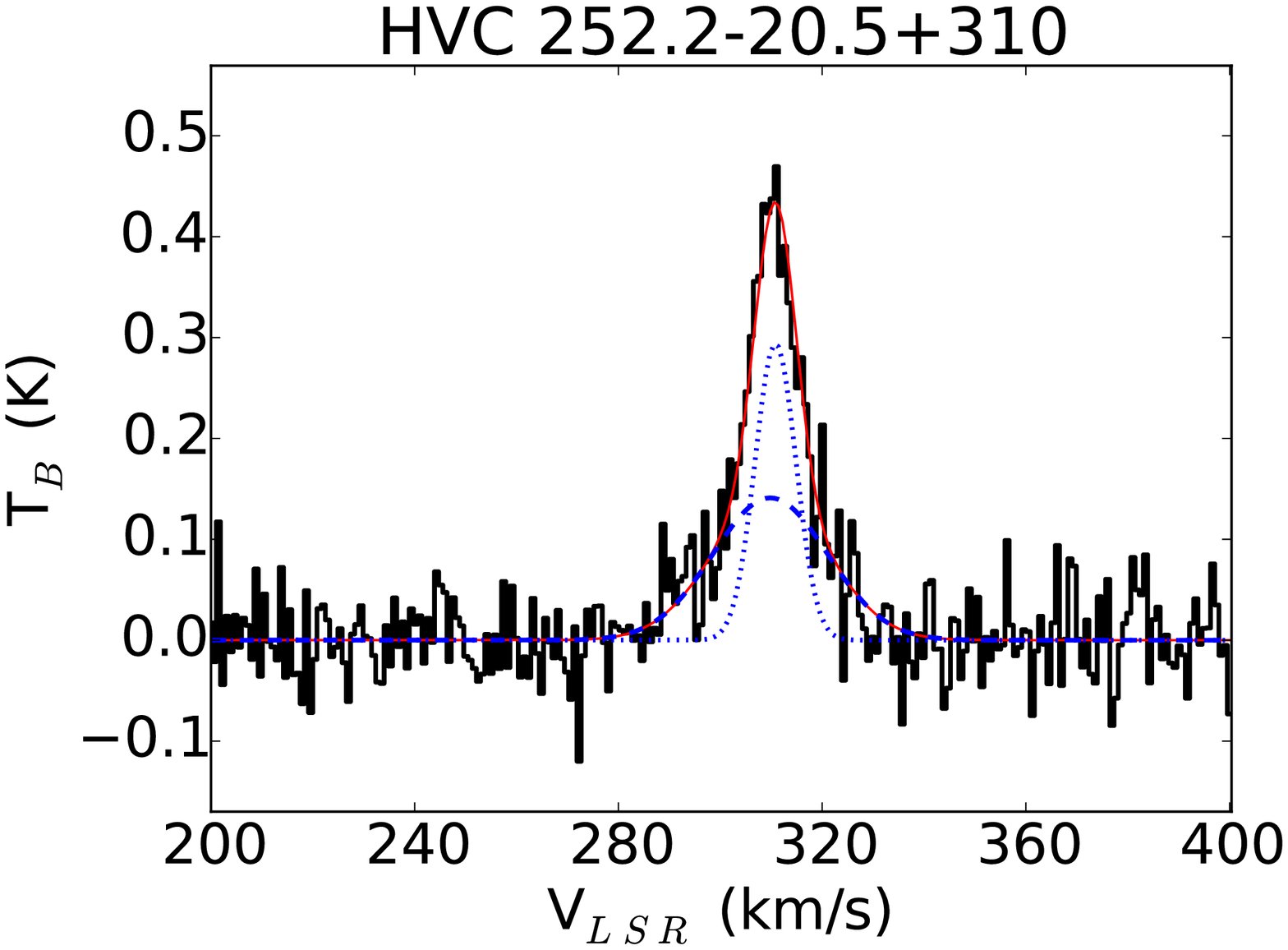}
\includegraphics[scale=0.4]{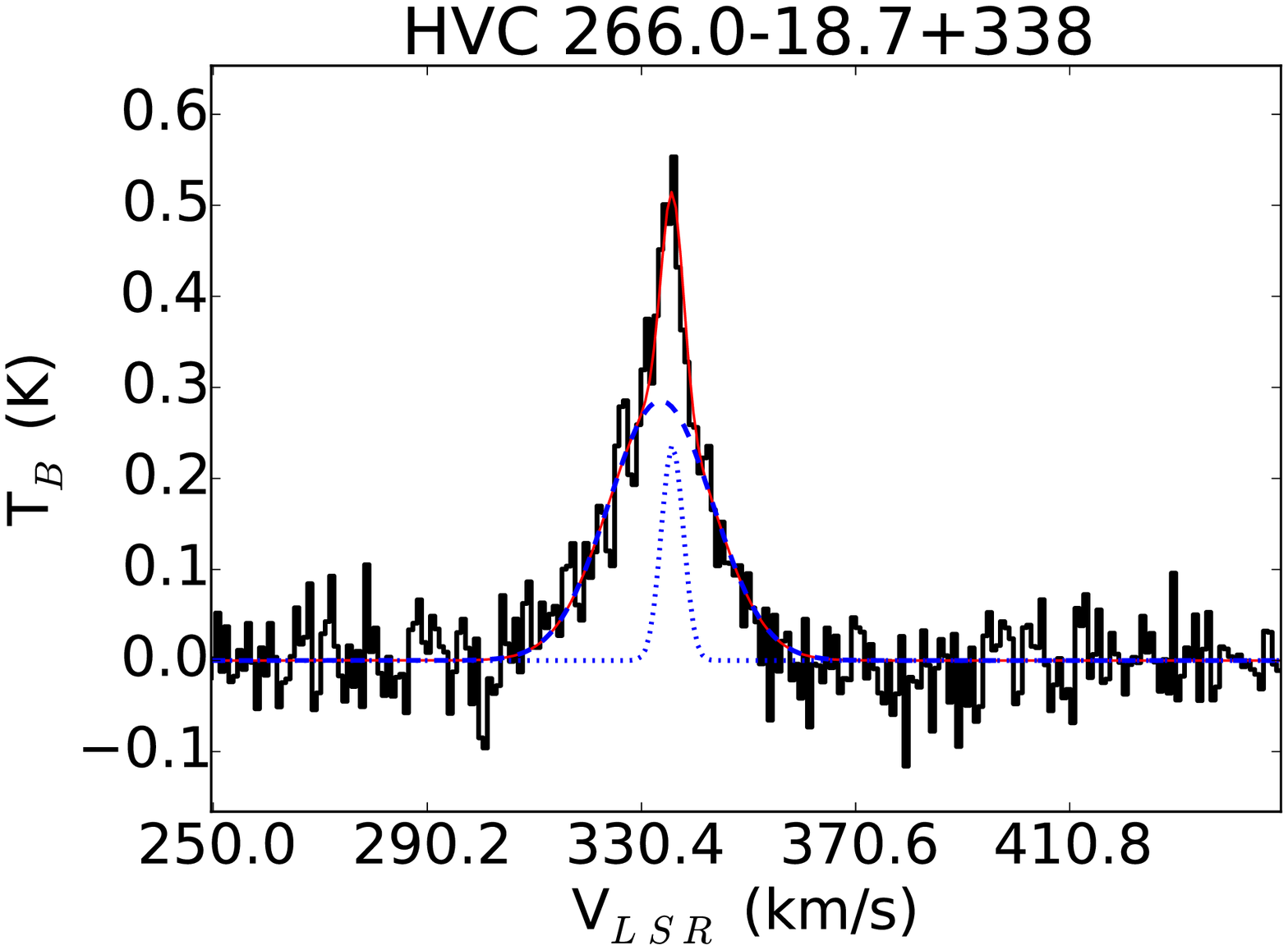}
\includegraphics[scale=0.4]{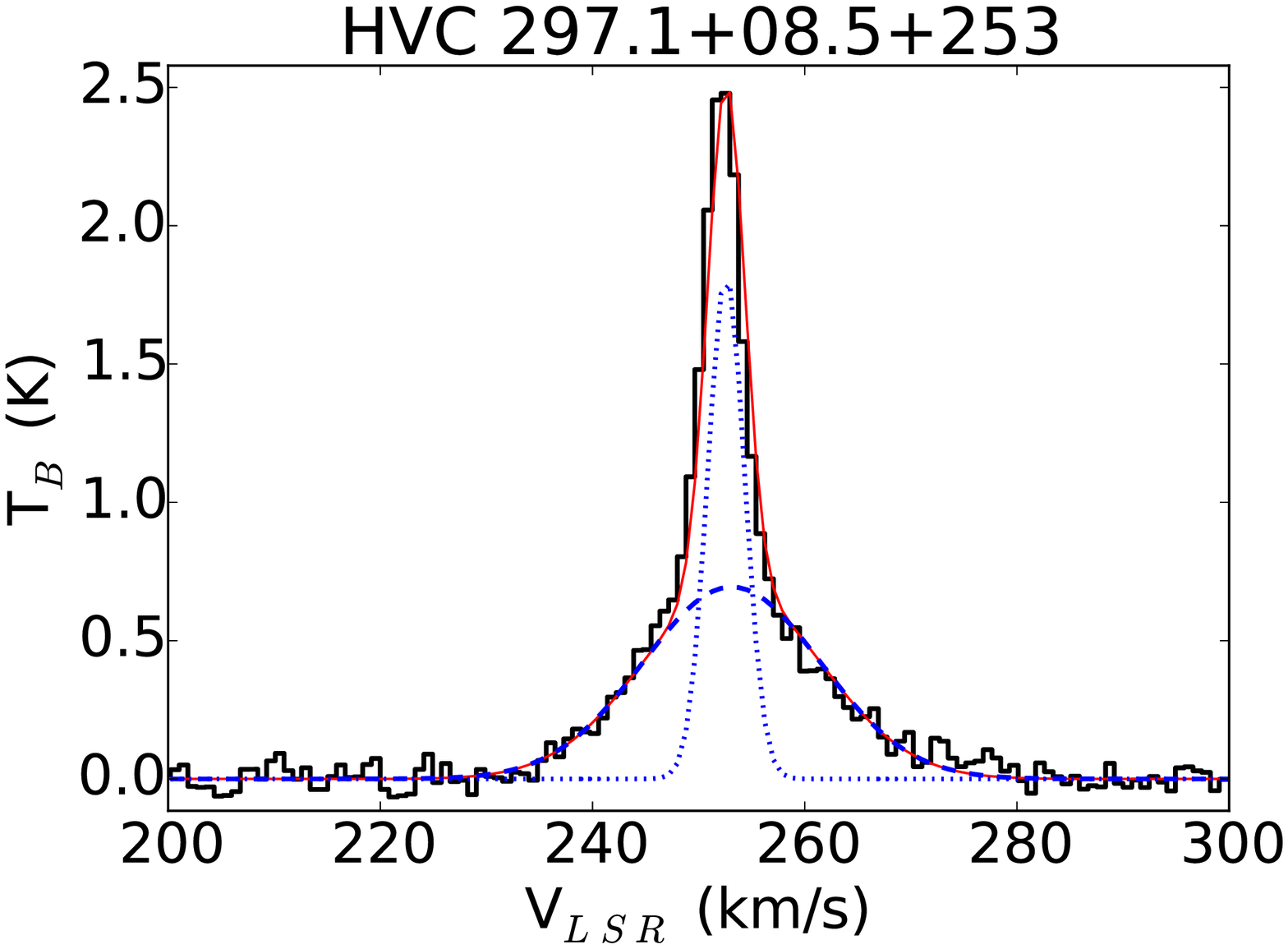}
\includegraphics[scale=0.4]{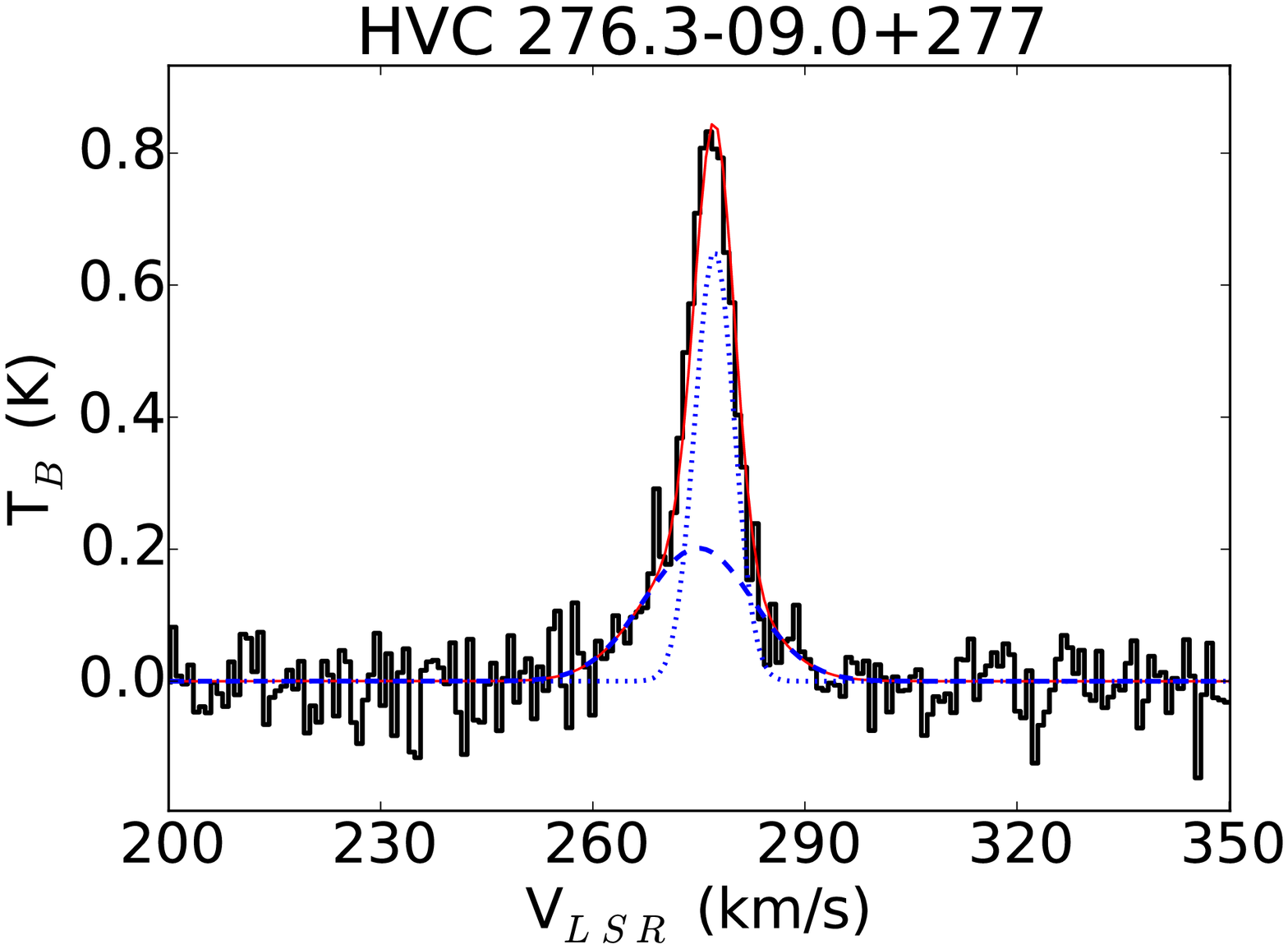}
\includegraphics[scale=0.4]{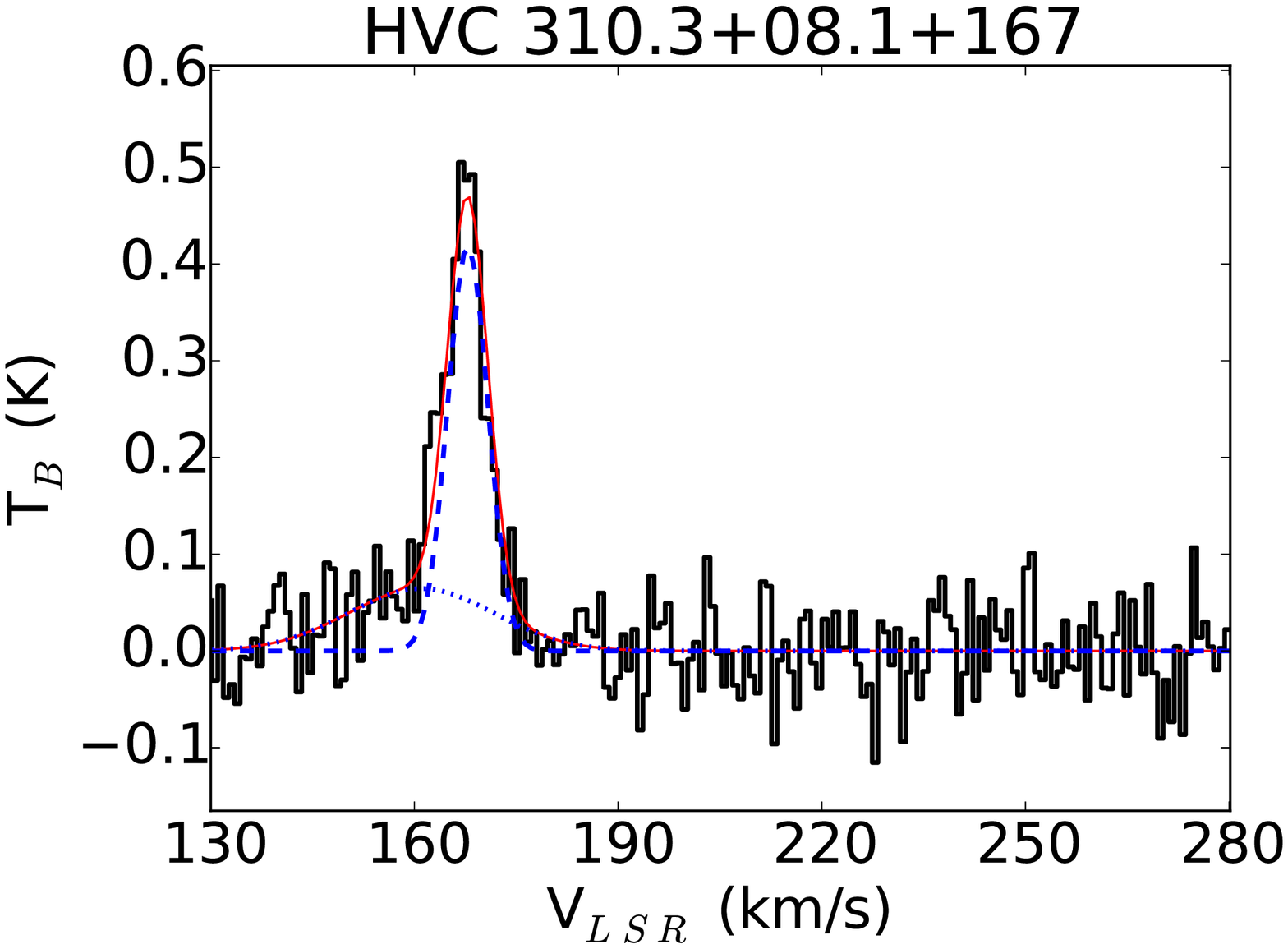}
\caption{Examples of double Gaussian fits to a spectrum from each cloud. 
Red indicates the combined profile of the two Gaussian profiles.}
\label{dgauss}
\end{figure*}

We analyze the GASS and the combined ATCA and GASS data separately for each cloud. 
For the GASS data, spectra are extracted along the symmetry axis of each cloud 
to derive the physical parameters of peak \HI\ column density ($N_{\rm HI}$), velocity in the 
Local Standard of Rest frame ($V_{\rm LSR}$) and velocity linewidth ($\Delta v$). 
In Figures~\ref{Gcomb1}--\ref{Gcomb5}, we show the extracted spectra (left), integrated 
\HI\ column density map (top right) and the derived physical parameters along the symmetry axis (bottom right) for each cloud. 
The white crosses in the integrated \HI\ column density map  
represent the positions of individual spectra along the sliced axis. 
The series of extracted spectra are fitted with either a single or 
double Gaussian profile, in which 
amplitude, velocity center and variance are set as free parameters. 
Examples of double Gaussian fits are shown in Figure~\ref{dgauss}.  
We calculate the integrated \HI\ column density using 
$N_{\rm HI}=1.823\times10^{18}\int T_{\rm B} dv= 1.823\times10^{18}\cdot T_{\rm B} \cdot \sigma\sqrt{2\pi}$, 
where $T_{\rm B}$ is peak brightness temperature in K, $v$ is the velocity in \kms, 
and $\sigma$ is the standard deviation in \kms. 
The velocity linewidth is calculated  
using the derived $\sigma$, $\Delta v$ = $\sigma \sqrt{8\ln(2)}$.   
The derived results are shown in the bottom right panel of Figures~\ref{Gcomb1}--\ref{Gcomb5}, 
with red crosses indicating 
the narrow linewidth component of the cloud. The derived physical parameters using GASS data are 
summarized in Table~\ref{GASSpar}. 
 
\begin{figure*}
\center
\includegraphics[scale=0.43]{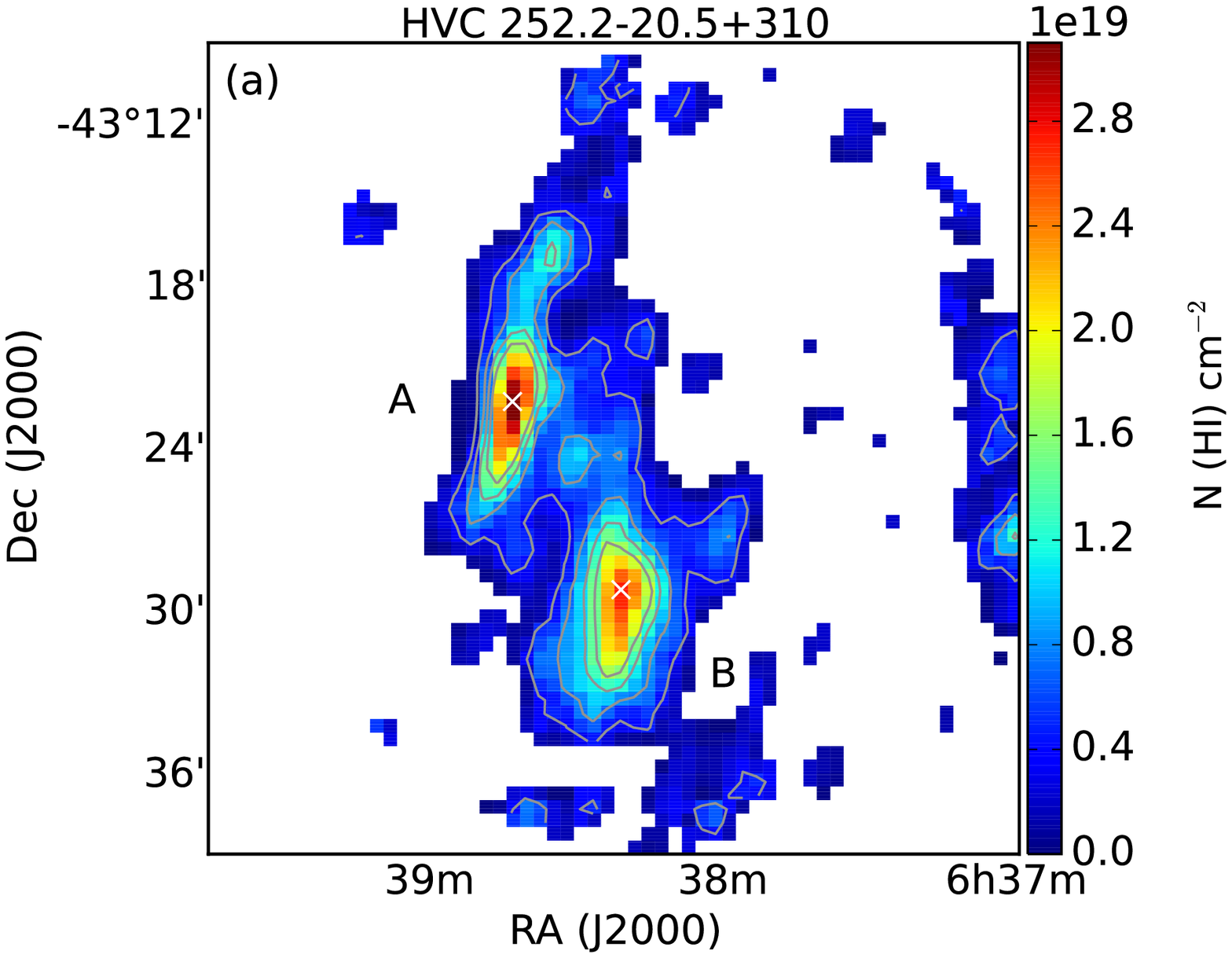}
\includegraphics[scale=0.43]{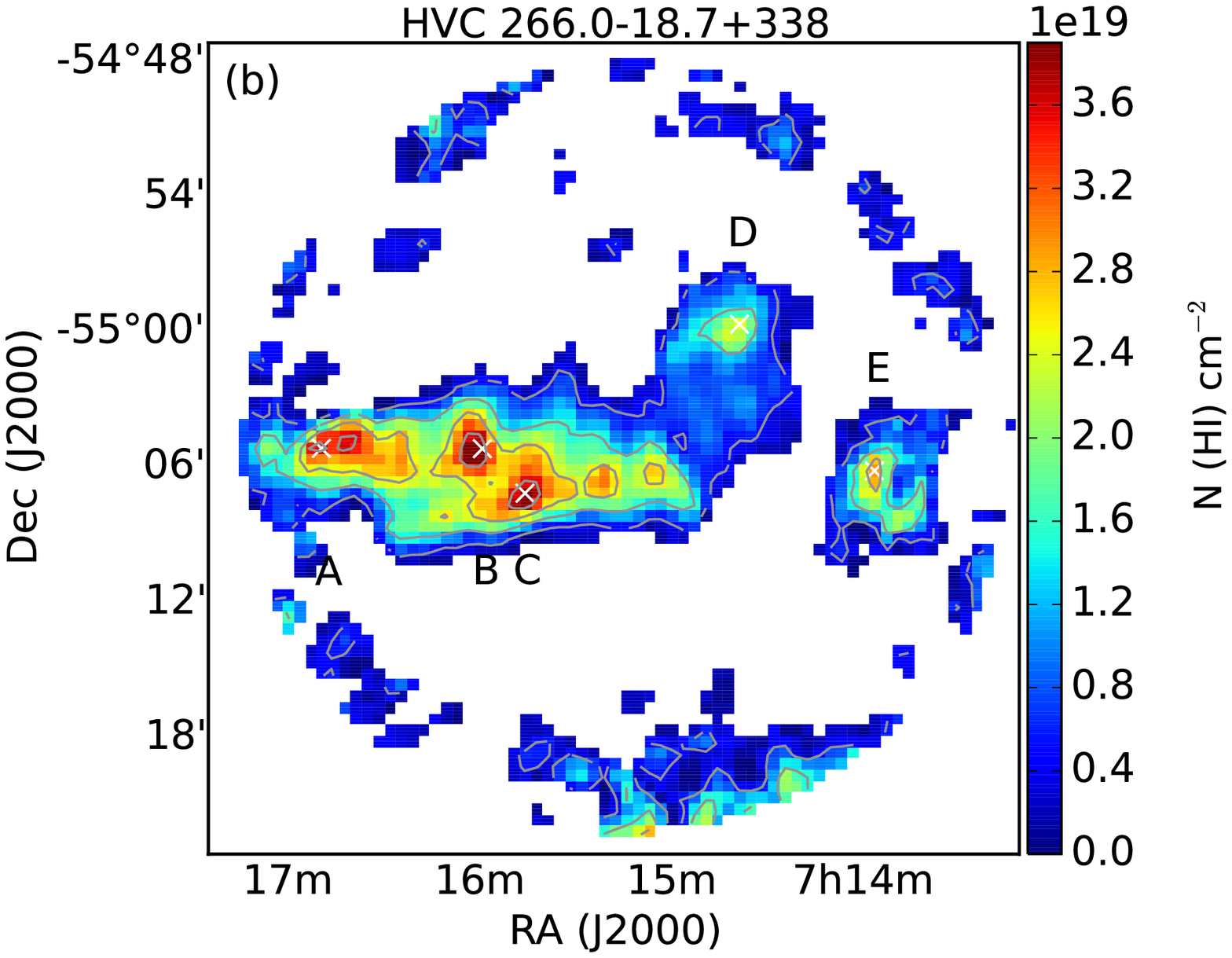}
\includegraphics[scale=0.42]{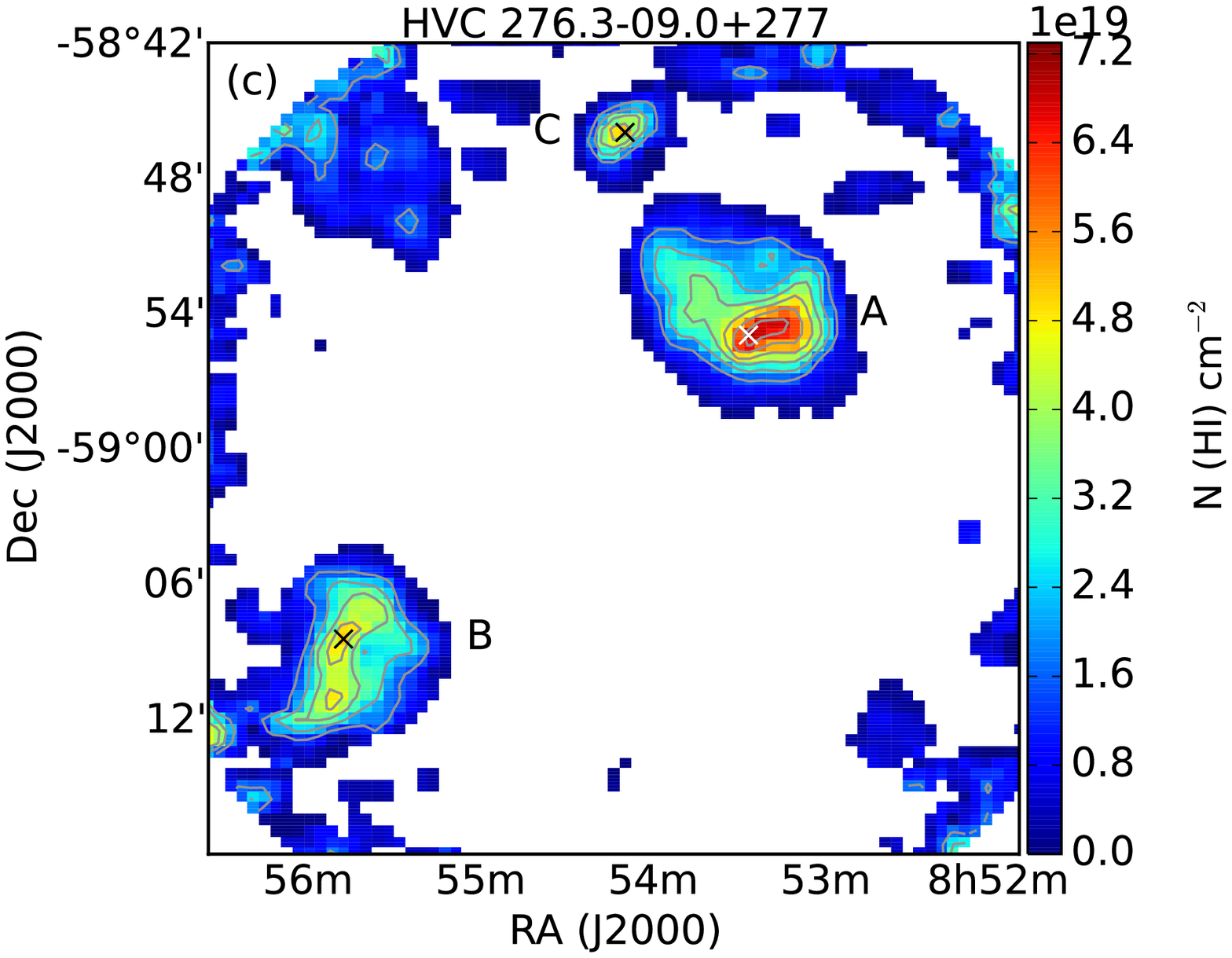}
\includegraphics[scale=0.42]{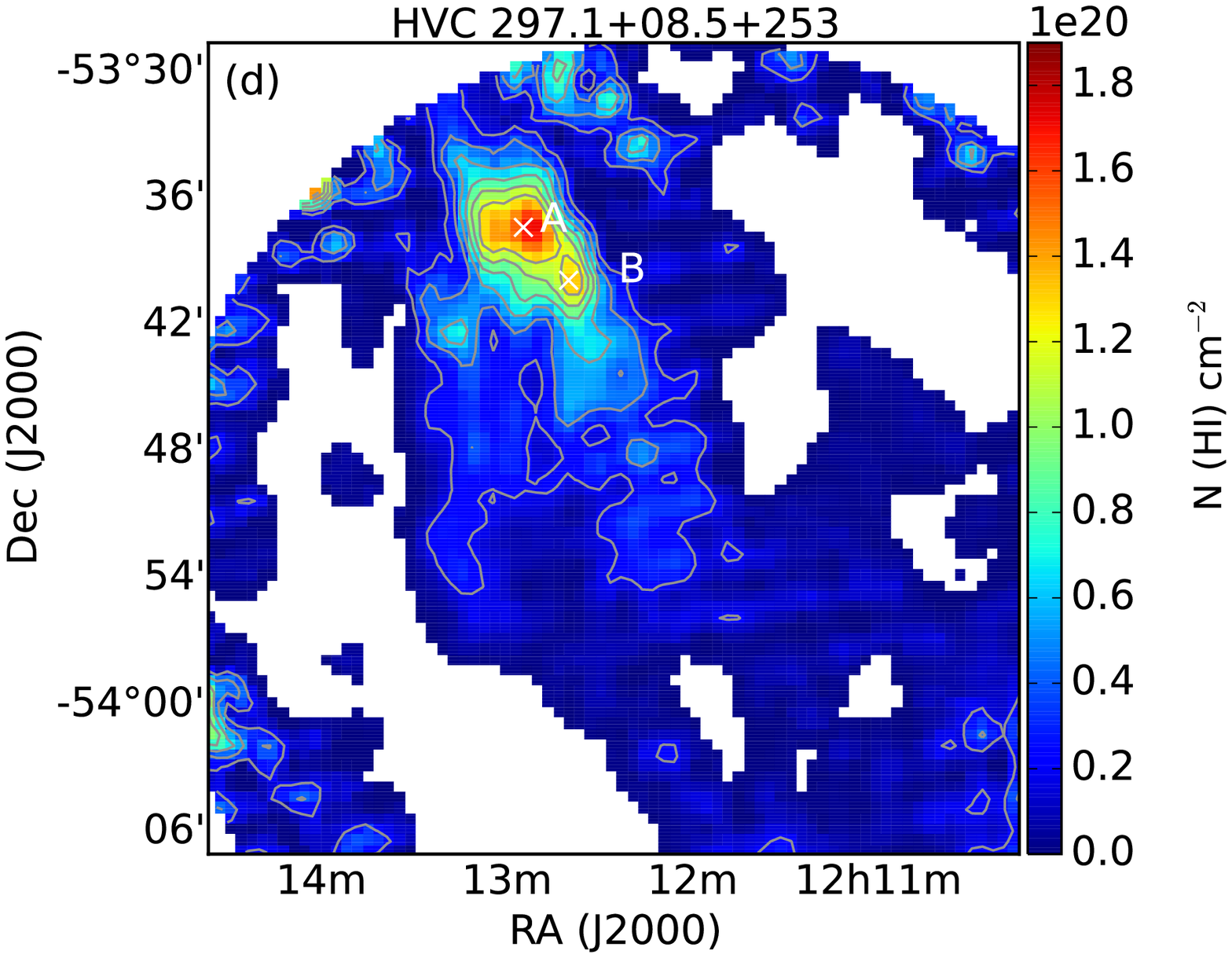}
\includegraphics[scale=0.42]{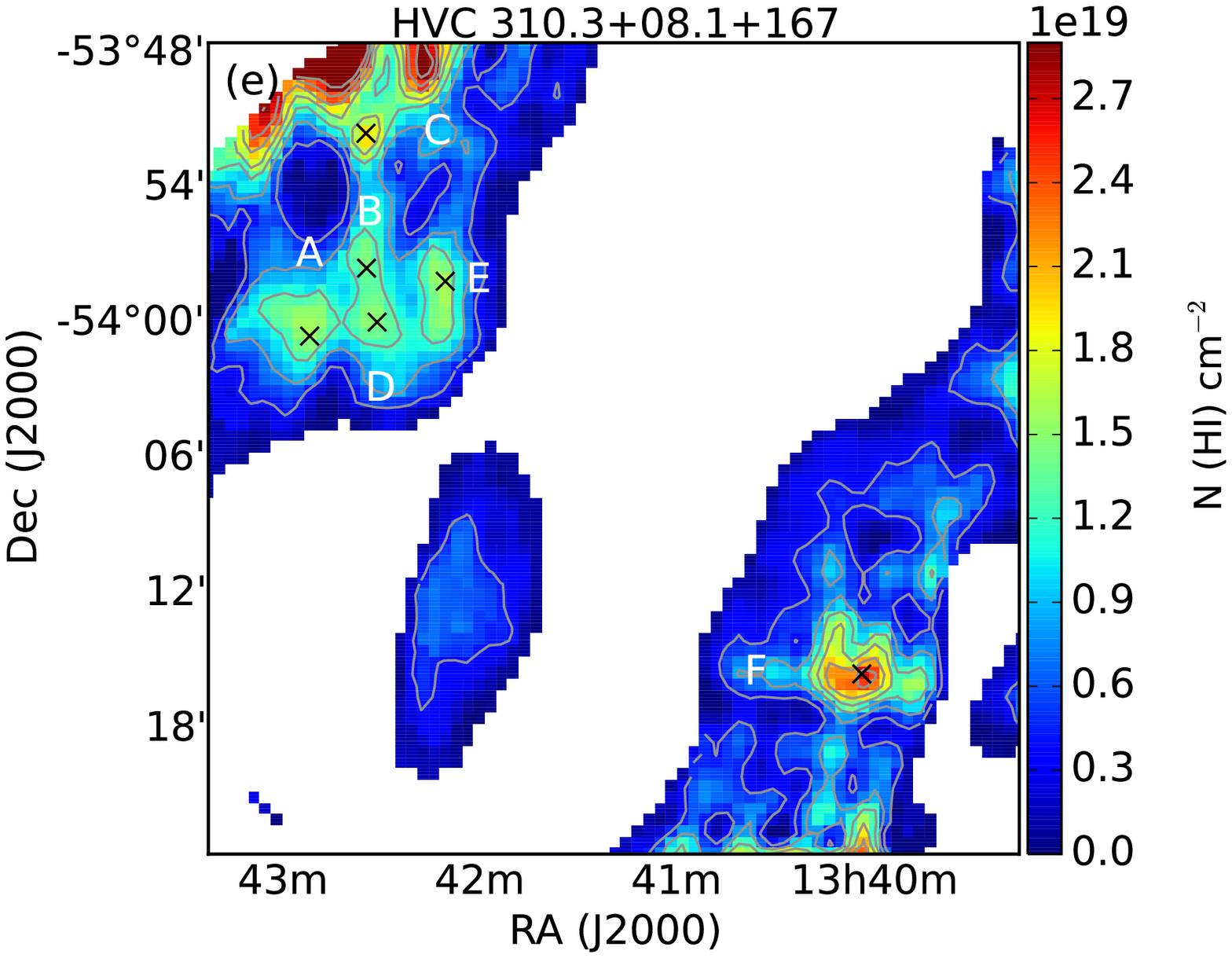}
\caption{Integrated \HI\ column density maps of the combined Parkes and 
ATCA \HI\ data sets. Identified clumps are marked with a cross and are labeled. 
Contours are plotted to highlight the morphology of the clumps. }
\label{clumps}
\end{figure*}

For the combined data, we run a 3-dimensional version of CLUMPFIND, which is an 
automatic routine for analyzing clumpy structure in spectral line data cubes \citep{Williams94}. 
The routine searches the local peaks of the emission and follows them down to the user defined 
intensity level. We note that 
CLUMPFIND has its limitations if the data cube contains significant background noise. 
After running CLUMPFIND, we inspect the output manually to eliminate false detections. 
Identified clumps in each cloud are shown in Figure~\ref{clumps} with crosses and labels. 
We perform a similar 
analysis as for the GASS data by extracting the spectra of the clumps at peak \HI\ column density 
and deriving their physical parameters. A summary of the derived physical parameters of these clumps 
is given in Table~\ref{clumps_tab}. 

\begin{table*}
\centering
\begin{minipage}{150mm}
\caption{Derived physical parameters of each cloud either using a single or double Gaussian profile. 
GASS data are used and the position number represents the location of extracted spectrum.}
\label{GASSpar}
\begin{tabular}{cccccccccc}
\hline
\hline
Name & 
Position & 
$T_{\rm B_{1}}$ & 
$T_{\rm B_{2}}$ & 
$V_{\rm LSR_{1}}$ &
$V_{\rm LSR_{2}}$ &
$N_{\rm HI_{1}}$ &
$N_{\rm HI_{2}}$ &
$\Delta v_{1}$ &
$\Delta v_{2}$ \\
& $\#$ & (K) & (K) & (\kms) & (\kms) & ($10^{18}$cm$^{-2}$) & ($10^{18}$cm$^{-2}$) & (\kms) & (\kms) \\
\hline
HVC 252.2-20.5+310	&	1	& \ldots	& 0.06 	& \ldots	& 306.1		& \ldots	& 3.30      	& \ldots	& 30.7		\\
                 	&	2	& \ldots      	& 0.10  & \ldots      	& 308.1 	& \ldots      	& 4.09      	& \ldots      	& 22.1  	\\
                        &	3	& \ldots      	& 0.18  & \ldots      	& 310.1 	& \ldots      	& 5.71      	& \ldots      	& 16.5  	\\
	&	4	& 0.22      	& 0.11         &	311.1	&	308.7	&	4.07	&	6.70	&	9.7	&	30.4	\\
	&	5	& 0.29      	& 0.14         &	310.8	&	309.8	&	5.38	&	7.24	&	9.4	&	26.5	\\
	&	6	& 0.26      	& 0.23         &	310.2	&	309.9	&	3.19	&	9.55	&	6.4	&	21.2	\\
	&	7	& 0.18      	& 0.23         &	310.6	&	309.2	&	1.67	&	6.98	&	4.8	&	15.6	\\
	&	8	& 0.20      	& 0.13         &	311.1	&	308.2	&	2.07	&	4.77	&	5.4	&	19.0	\\
	&	9	& 0.14      	& 0.10         &	310.4	&	305.3	&	2.34	&	6.57	&	8.5	&	35.5	\\
	&	10	& 0.08      	& 0.08         &	309.9	&	309.7	&	1.29	&	4.81	&	7.9	&	32.1	\\
	&	11	& 0.07      	& 0.07         &	310.7	&	309.6	&	0.63	&	3.87	&	4.7	&	28.3	\\
	&	12	& \ldots       	& 0.08         &	\ldots	&	307.8	&	\ldots  &	4.08	&	\ldots  &	24.9	\\
\hline			                               												
HVC 266.0-18.7+338	&	1	& 0.20      	& \ldots       &	340.5	&	\ldots	&	3.61	&	\ldots	&	9.4	&	\ldots	\\
	&	2	& 0.20      	& 0.21         &	339.6	&	341.5	&	1.63	&	4.71	&	4.2	&	11.3	\\
	&	3	& 0.30      	& 0.30         &	339.8	&	340.5	&	2.48	&	7.13	&	4.3	&	12.2	\\
	&	4	& 0.46      	& 0.26         &	339.8	&	339.0	&	4.68	&	8.94	&	5.3	&	18.0	\\
	&	5	& 0.48      	& 0.27         &	339.1	&	337.6	&	6.22	&	11.21	&	6.7	&	21.3	\\
	&	6	& 0.49      	& 0.22         &	338.1 	&	335.5	&	8.56	&	12.09	&	9.1	&	27.8	\\
	&	7	& 0.35      	& 0.24         &	336.9	&	334.7	&	6.11	&	12.49	&	9.0	&	27.3	\\
	&	8	& 0.24      	& 0.29         &	336.2	&	334.0	&	2.24	&	12.71	&	4.9	&	23.0	\\
	&	9	& 0.18      	& 0.23         &	336.1	&	333.4	&	1.34	&	9.51	&	3.9	&	21.3	\\
	&	10	& 0.13      	& 0.13         &	335.6	&	332.8	&	1.35	&	6.06	&	5.5	&	23.7	\\
	&	11	& \ldots      	& 0.13         &	\ldots 	&	333.7	&	\ldots	&	4.68	&	\ldots	&	18.3	\\
\hline			                               												
HVC 276.03-09.0+277	&	1	& \ldots      	& 0.16       &	\ldots	&	277.6	&	\ldots	&	4.64	&  \ldots	&	14.6	\\
                	&	2	& 0.18        	& 0.20       &	280.0 	&	277.4   &	2.21  	&	6.84    &  6.5	       	&	17.4      \\
                 	&	3	& 0.37        	& 0.09       &	279.5 	&	272.4	&	7.10  	&	2.80	&  10.0       	&	16.2	\\
	&	4	& 0.27      	& \ldots       &	278.0	&	\ldots	&	5.20	&	\ldots	&	9.9	&	\ldots	\\
	&	5	& 0.45     	& 0.07         &	277.2	&	273.4	&	6.27 	&	2.59	&	7.1	&	19.6	\\
	&	6	& 0.65      	& 0.20         &	277.2	&	274.9	&	8.26	&	7.12	&	6.5	&	18.2	\\
	&	7	& 0.65      	& 0.10         &	277.1	&	274.4	&	8.80	&	4.13	&	7.0	&	21.3	\\
	&	8	& 0.30      	& \ldots       &	276.8	&	\ldots	&	4.52	&	\ldots	&	7.7	&	\ldots	\\
\hline			                               												
HVC 297.1+08.5+253	&	1	& \ldots   	&  0.17         &	\ldots 	&	249.7 &	\ldots  &	 7.31  & \ldots		&	22.1	\\
                  	&	2	& \ldots   	&  0.30         &	\ldots 	&	251.1 &	\ldots	&	 13.30 & \ldots       	&	23.0	\\
                        &	3	& \ldots   	&  0.44         &	\ldots 	&	253.1 &	\ldots	&	 19.92 & \ldots       	&	23.4	\\
                   	&	4	& \ldots   	&  0.47         &	\ldots 	&	253.8 &	\ldots	&	 20.05 & \ldots       	&	21.8	\\
	                &	5	& \ldots   	&  0.57         &	\ldots 	&	254.0 &	\ldots	&	 24.23 & \ldots       	&	22.0	\\
                      	&	6	& \ldots   	&  0.64         &	\ldots 	&	253.3 &	\ldots	&	 26.46 & \ldots       	&	21.4	\\
                 	&	7	& 0.21     	&  0.70         &	252.9  	&	252.5 &	2.55  	&	 29.74 & 6.1          	&	22.0  \\
                 	&	8	& 0.46     	&  0.76         &	252.7  	&	252.9 &	3.81  	&	 31.90 & 4.3          	&	21.6  \\
                	&	9	& 1.15     	&  0.67         &	252.6  	&	252.9 &	8.69  	&	 20.31 & 3.9          	&	15.5  \\
                	&	10	& 1.82     	&  0.69         &	252.6  	&	253.1 &	14.52 	&	 26.57 & 4.1          	&	19.7  \\
	&	11	& 1.99      	& 0.54         &	252.6	&	252.9	&	15.28	&	17.92	&	4.0	&	17.3	\\
	&	12	& 2.02      	&  0.38        &	252.6	&	252.7	&	12.92	&	13.54	&	3.3	&	18.4	\\
	&	13	& 1.03      	&  0.22        &	252.5	&	252.5	&	6.45 	&	5.87 	&	3.2	&	13.6	\\
	&	14	& 0.35      	& 0.27         &	252.0	&	252.9	&	1.62	&	3.05	&	2.4	&	5.8	\\
\hline			                               												
HVC 310.3+08.1+167	&	1	& 0.07         	& 0.08     &	161.3 	& 160.0	  &	0.46  	&	1.99    &	3.4	&	13.2    \\
	                &	2	& \ldots       	& 0.14     &	\ldots	& 160.7	  &	\ldots	&	3.39	&	\ldots	&	12.2	\\
                  	&	3	& \ldots       	& 0.17     &	\ldots	& 161.1	  &	\ldots	&	6.93	&	\ldots	&	21.3	\\
                  	&	4	& 0.06      	& 0.12         &	160.1	&	163.6	&	0.86	&	4.95	&	7.0	&	21.3	\\

                 	&	5	&  \ldots   	&  0.15        & \ldots		&	163.7	&	\ldots 	& 4.07	 & 	\ldots		&14.4	\\
                	&	6	&  \ldots   	&  0.16        & \ldots       	&	165.0	&	\ldots 	& 4.56   & 	\ldots     	&14.5	\\
                    	&	7	&  0.11     	&  0.14        & 164.7        	&	166.5   &	1.77   	& 4.97   & 	8.6        	&18.2  \\
                  	&	8	&  0.10     	&  0.12        & 166.7        	&	165.9   &	0.79   	& 4.41   & 	4.0        	&18.4  \\
                   	&	9	&  \ldots   	&  0.20        & \ldots       	&	166.6	&	\ldots 	& 4.98   & 	\ldots     	&13.1	\\
                 	&	10	&  \ldots   	&  0.36        & \ldots       	&	168.3	&	\ldots 	& 7.05   & 	\ldots     	&10.1	\\
                 	&	11	&  0.17     	&  0.42        & 168.9        	&	168.1	&	0.67   	& 7.63 	 & 	2.1        	&9.4 	\\
                    	&	12	& 0.42      	& 0.06         &	167.9	&	161.0	&	5.69	&	3.16	&	7.0	&	25.3	\\
                  	&	13	& 0.20      	& 0.08         &	168.3	&	164.0	&	2.36	&	2.38	&	6.1	&	15.3	\\
\hline
\end{tabular}
\end{minipage}
\end{table*}

\begin{sidewaystable}
\tiny
\centering
\begin{minipage}{250mm}
\caption{Derived physical parameters and location of each clump found in the combined data.}
\label{clumps_tab}
\begin{tabular}{cccccccccccccccc}
\hline
\hline
Name & 
Clump & 
RA & 
Dec & 
$T_{\rm B_{1}}$ &
$T_{\rm B_{2}}$ &
$V_{\rm LSR_{1}}$ &
$V_{\rm LSR_{2}}$ &
$N_{\rm HI_{1}}$ &
$N_{\rm HI_{2}}$ &
$\Delta v_{1}$ &
$\Delta v_{2}$ &
$T_{\rm k}$ &
$\theta$ &
$n^{b}$ &
$P/k_{\rm B}^{b}$ 
\\
 &  & & & (K) & (K) & (\kms) & (\kms) & ($10^{19}$~cm$^{-2}$) &  ($10^{19}$~cm$^{-2}$) & (\kms) & (\kms) & (K) & ($\degr$) & (cm$^{-3}$) & (K cm$^{-3}$)\\
\hline
HVC 252.2-20.5+310	&	A	&	06$^{h}$38$^{m}$42.26$^{s}$	&	$-43\degr22\arcmin14.43\arcsec$	&	3.49	&	1.27	&	312.5	&	313.4	&	1.89	&	1.82	&	2.80	&	7.38	& \ldots	&	\ldots	&	\ldots	&	\ldots  	\\
	                &	B	&	06$^{h}$38$^{m}$20.27$^{s}$	&	$-43\degr29\arcmin14.73\arcsec$	&	2.78	&	\ldots	&	310.8	&	\ldots	&	2.50	&	\ldots	&	4.65	&	\ldots	& 471.37	&	0.144	&	0.13	&	60.78         \\
\hline	                                                                                                                                                                                                                                                  	
HVC 266.0-18.7+338	&	A	&	07$^{h}$16$^{m}$36.77$^{s}$	&	$-55\degr05\arcmin14.92\arcsec$	&	3.29	&	\ldots	&	340.9	&	\ldots	&	4.04	&	\ldots	&	6.33	&	\ldots	& 872.96	&	0.052	&	0.58	&	503.19        \\
	                &	B	&	07$^{h}$16$^{m}$01.82$^{s}$	&	$-55\degr05\arcmin16.69\arcsec$	&	3.78	&	\ldots	&	341.3	&	\ldots	&	4.34	&	\ldots	&	5.92	&	\ldots	& 764.99	&	0.040	&	0.80	&	610.84        \\
	                &	C	&	07$^{h}$15$^{m}$44.38$^{s}$	&	$-55\degr07\arcmin17.20\arcsec$	&	3.52	&	\ldots	&	339.5	&	\ldots	&	4.37	&	\ldots	&	6.40	&	\ldots	& 893.88	&	0.048	&	0.68	&	604.95        \\
	                &	D	&	07$^{h}$14$^{m}$38.03$^{s}$	&	$-54\degr59\arcmin46.95\arcsec$	&	1.24	&	\ldots	&	338.2	&	\ldots	&	2.51	&	\ldots	&	10.41	&	\ldots	& 2361.77	&	0.056	&	0.33	&	779.77        \\
	                &	E	&	07$^{h}$13$^{m}$55.98$^{s}$	&	$-55\degr06\arcmin15.34\arcsec$	&	2.07	&	\ldots	&	336.0	&	\ldots	&	5.28	&	\ldots	&	13.16	&	\ldots	& 3772.62	&	0.049	&	0.80	&	3017.05       \\
\hline	                                                                                                                                                                                                                                                  	
HVC 276.03-09.0+277	&	A	&	08$^{h}$53$^{m}$26.09$^{s}$	&	$-58\degr54\arcmin57.38\arcsec$	&	6.67	&	\ldots	&	276.5	&	\ldots	&	7.95	&	\ldots	&	6.14	&	\ldots	& 823.07	&	0.056	&	1.05	&	864.63        \\
	                &	B	&	08$^{h}$55$^{m}$46.20$^{s}$	&	$-59\degr08\arcmin54.81\arcsec$	&	3.69	&	\ldots	&	281.0	&	\ldots	&	4.70	&	\ldots	&	6.56	&	\ldots	& 937.18	&	0.042	&	0.83	&	781.87        \\
	                &	C	&	08$^{h}$54$^{m}$08.70$^{s}$	&	$-58\degr45\arcmin56.93\arcsec$	&	2.39	&	5.99	&	273.5	&	275.5	&	4.39	&	3.78	&	9.44	&	3.25	& \ldots	&	\ldots	&	\ldots	&	\ldots        \\
\hline	                                                                                                                                                                                                                                                  	
HVC 297.1+08.5+253	&	A	&	12$^{h}$12$^{m}$51.88$^{s}$	&	$-53\degr37\arcmin28.63\arcsec$	&	20.29	&	4.38	&	252.8	&	249.5	&	12.99	&	1.88	&	3.30	&	2.21	& \ldots	&	\ldots	&	\ldots	&	\ldots        \\
	                &	B	&	12$^{h}$12$^{m}$38.40$^{s}$	&	$-53\degr39\arcmin29.49\arcsec$	&	11.69	&	\ldots	&	252.9	&	\ldots	&	11.13	&	\ldots	&	4.91	&	\ldots	& 524.91	&	0.028	&	3.00	&	1572.20       \\
\hline	                                                                                                                                                                                                                                                  	
HVC 310.3+08.1+167$^{a}$	&	A	&	13$^{h}$42$^{m}$50.09$^{s}$	&	$-54\degr00\arcmin33.46\arcsec$	&	0.72	&	0.23	&	165.5	&	172.3	&	0.70	&	0.15	&	5.02	&	3.31	& \ldots	&	\ldots	&	\ldots	&	\ldots        \\
	                &	B	&	13$^{h}$42$^{m}$32.98$^{s}$	&	$-53\degr57\arcmin34.32\arcsec$	&	0.30	&	0.61	&	172.2	&	166.2	&	0.21	&	0.70	&	3.56	&	5.94	& \ldots	&	\ldots	&	\ldots	&	\ldots        \\
	                &	C	&	13$^{h}$42$^{m}$32.80$^{s}$	&	$-53\degr51\arcmin33.45\arcsec$	&	0.29	&	0.57	&	163.5	&	169.5	&	0.26	&	0.59	&	4.69	&	5.34	& \ldots	&	\ldots	&	\ldots	&	\ldots        \\
	                &	D	&	13$^{h}$42$^{m}$29.65$^{s}$	&	$-54\degr00\arcmin04.75\arcsec$	&	0.29	&	0.55	&	172.5	&	166.6	&	0.20	&	0.63	&	3.42	&	5.86	& \ldots	&	\ldots	&	\ldots	&	\ldots        \\
	                &	E	&	13$^{h}$42$^{m}$09.20$^{s}$	&	$-53\degr58\arcmin05.56\arcsec$	&	0.32	&	0.12	&	171.9	&	167.1	&	0.28	&	0.11	&	4.56	&	4.63	& \ldots	&	\ldots	&	\ldots	&	\ldots        \\
	                &	F	&	13$^{h}$40$^{m}$02.87$^{s}$	&	$-54\degr15\arcmin34.20\arcsec$	&	0.28	&	0.16	&	167.4	&	161.6	&	0.31	&	0.17	&	5.57	&	5.47	& \ldots	&	\ldots	&	\ldots	&	\ldots        \\
\hline
\end{tabular}

$a$: Parameters are derived from a spatially and spectrally smoothed data cube.\\
$b$: Assuming a distance of 25~kpc.

\end{minipage}
\end{sidewaystable}

\section{OVERALL PROPERTIES OF THE HVCS}

Four of the HVCs studied here are classified as head-tail clouds. 
The only exception, HVC~266.0$-$18.7+338, is classified 
as a symmetric cloud by FSM13. 
However, the outer diffuse envelope makes the classification ambigious. 
It may also be a head-tail cloud. 

The head-tail morphology suggests ram-pressure interaction with the ambient 
medium. FSM13 reveal multiphase structures, which resemble broad and narrow components. 
We consider the broad component to be $>$10~\kms\ in this study. 
On average, the velocity linewidths for the broad and narrow component are 20~\kms\ and 
6.2~\kms, respectively. Most of the identified clumps only have a cold core. There are 
also unresolved clumps showing multiphase structures with broad and cold components or 
cold components only. 

\subsection{HVC 252.2$-$20.5+310}

HVC~252.2-20.5+310 is part of the LA IV. The tail is 
pointing away from the Galactic Plane. In Figure~\ref{Gcomb1},
we show the spectral line profiles and derived physical parameters 
along the sliced axis of the cloud.  
The cloud is sliced from south to north with a designated 
position number.
Multiphase structures of warm (broad) and cold (narrow) components at the 
head of the cloud are seen in the series of spectral line profiles. 
The rise and fall of the \HI\ column 
density is gradual. There is no obvious velocity gradient 
across this cloud. We find that the cold component (red crosses) 
has a fairly consistent 
$V_{\rm LSR}\sim311$~\kms. The FWHM of the warm component (black circles) 
varies in different parts of the cloud. The cold component has 
a typical linewidth of 7~\kms.

Figure~\ref{merged_GASScon}a shows the peak \HI\ column density map of the 
combined image. The overlaid contour provides the spatial size of the cloud as observed by GASS. 
As seen in the figure, the 
head of the cloud has been resolved into two clumps. The morphology of these two clumps 
is outlined with contours in Figure~\ref{clumps}a. Clumps A and B morphologically 
look like head-tail clouds which are pointing in the opposite direction. 
The derived physical parameters at the peak \HI\ column density of both clumps 
are presented in Table~\ref{clumps_tab}. 
Clump A has two unresolved components, which have $\Delta v$ of 2.80 and 7.38~\kms.  
Clump B is resolved with a single peak $T_{\rm B}$ of 2.78~K. 

\subsection{HVC 266.0$-$18.7+338}

HVC~266.0-18.7+338 is also part of the LA IV. It is classified 
as a symmetric cloud in FSM13. The cloud is 
sliced from east to west, which corresponds to the 
spectral line profiles from top to bottom in Figure~\ref{Gcomb2}. 
The cloud shows a clear velocity gradient with velocity decreasing from 
about 340~\kms\ to 332~\kms. In general, the cold component has a velocity 
larger than the warm component. The effect can be seen in the 
series of asymmetric spectral line profiles. The $\Delta v$ of the warm component 
increases from the east end of the cloud and reaches a maximum at the center 
before flattening it out to $\sim25$~\kms. A similar trend is seen in 
$\Delta v$ of the cold component. The linewidth and velocity gradients indicate 
that this cloud is probably a head-tail cloud rather than a symmetric cloud. 

The ATCA primary beam almost covers the entire cloud (see Figure~\ref{merged_GASScon}b). 
The combined image reveals a complex structure. 
Figure~\ref{clumps} shows a total of 5 distinct clumps identified via 
CLUMPFIND, named A--E. All clumps are resolved. These clumps have a 
fairly similar $V_{\rm LSR}$ (336--341~\kms). Clump D and E 
are considered warm with $\Delta v >$ 10~\kms. 
Both of them are also isolated from the main concentration of other clumps. 
None of them has a distinct morphology. 
  
\subsection{HVC 276.3$-$09.0+277}

HVC 276.3-09.1+277 is the third HVC in this study that is located in LA IV. 
This head-tail cloud has two 
main cores with one being at the head of the cloud and the other at the tail 
(see Figure~\ref{Gcomb3}). The analysis is performed by slicing the cloud 
diagonally from south-east to north-west. 
The cloud does not show a clear velocity gradient. 
Its head is rather compressed. 
The average $\Delta v$ for all of the cold components 
is 7.5~\kms. The $V_{\rm LSR}$ values of the cold components decrease gradually and 
are larger than the warm components. The peak $T_{\rm B}$ is $\sim$0.7~K at the main
core of the cloud. 

In Figures~\ref{merged_GASScon}c and \ref{clumps}c, 
we show the resolved clumps in the combined image. Three clumps are identified. 
Clump A is located in the main core of the cloud 
but clump B is displaced slightly from the second core of the cloud as seen 
in the diffuse emission. 
Both clump A and B are morphologically similar to head-tail clouds. 
Clump C is relatively smaller than clump A and B. It is unresolved 
and consists of two cold components, which have $\Delta v$ of 3.25 
and 9.44~\kms. 

\subsection{HVC 297.1+08.5+253}

HVC~297.1+08.5+253 is situated in the vicinity between LA I and II, 
closer to the south-eastern part of LA II. This head-tail cloud 
is morphologically slightly different from HVC 252.2$-$20.5+210. The 
contours show a typical head-tail structure (see Figure~\ref{Gcomb4}).  
The tail of HVC~297.1+08.5+253 is pointing away 
from the Galactic plane. 

In Figure~\ref{Gcomb4}, the 
spectral line profiles (top to bottom) represent the sliced positions 
of the cloud from south-west to north-east. As expected, 
the peak \HI\ column density increases toward the head of the cloud. 
Interestingly, the derived \HI\ column density of the warm component 
reaches its maximum (position 8) before the 
maximum \HI\ column densities as shown in the GASS integrated 
\HI\ column density map at positions 9 and 10. The cold component 
also shows the same rise and fall pattern in \HI\ column density but 
with the maximum at position 11. The $V_{\rm LSR}$ shows a small gradient 
at the tail of the cloud and then becomes constant at 253~\kms\ for both 
cold and warm component. Half of the cloud consists of warm component 
only, with an average velocity linewidth of 20.1~\kms. 
At position 14, two components are detected (refer bottom panel of Figure~6). 
These two components on avarage have a velocity linewidth of 4.1~\kms.    

The cloud is relatively diffuse as shown in the combined image 
(Figure~\ref{merged_GASScon}d). The main resolved feature is 
offset and leading the diffuse core of the cloud as seen in the GASS 
image. This explains why the \HI\ column density 
of the cold component peaks at position 11 in the analysis of the GASS 
image. Figure~\ref{clumps}d shows the resolved 
feature and two detected clumps. Unfortunately, the feature is 
detected fairly close to the edge of the ATCA beam, where sensitivity 
drops off significantly. Thus, extended features cannot be seen.

Clump A has the highest $T_{\rm B}$ ($\sim$20.3~K) 
amongst the detected clumps in this study. Clump A is resolved and 
reveals sub-components as seen in its spectral line profile (not shown here). 
The resolved clump B has $\Delta v$ of $\sim5$~\kms.      

\subsection{HVC 310.3+08.1+167}

HVC 310.3+08.1+167 is located north of LA I. 
It is pointing in the general direction of motion of LA I. This cloud 
is quite diffuse compared to the other clouds in this study, 
It has a very long diffuse tail, which spans nearly 1.5$\degr$, and a 
slight kink at the end (see Figure~\ref{Gcomb5}). While 
this cloud has a lower $V_{\rm LSR}$ than other selected HVCs in this study, 
it is assumed to be part of the LA based on the constraints listed in FSM13 to 
exclude the Galactic \HI\ emission. 

The analysis is performed by slicing along the symmetry axis from 
south-west to north-east. The spectral line profiles show that the cloud has 
a fairly low brightness temperature of $\sim$0.4~K. 
The multiple peaks of \HI\ column density indicate numerous components. 
Apart from the two deviating points at position 12 and 13 in the velocity 
plot, the gradual 
increase of $V_{\rm LSR}$ suggests that the cloud has a velocity gradient. 

The contours in Figure~\ref{merged_GASScon}e show that 
the size of the cloud is a lot larger than the ATCA beam. 
Since its coordinates were derived at the axis center of the cloud, 
the core structure was not observed at the center of the ATCA beam, where 
sensitivity is highest. The low surface brightness nature of 
the cloud also makes clump idenfication harder. 
The identification and derivation of physical parameters of the clumps were performed 
on the smoothed cube. Six distinct clumps 
have been identified (see Figure~\ref{clumps}e). 
All of them are unresolved and have sub-components. The sub-components are cold 
as well. 

The structure near the head is complex with five clumps being detected 
within the sensitivity cutoff. They have $V_{\rm LSR}$ ranging 
from 165 to 172~\kms. Clump F has a lower $V_{\rm LSR}$ 
of 161.6~\kms. They all appear to have an irregular shape.    

\section{COMPARISON WITH PREVIOUS STUDIES\label{comp}}

Two additional compact HVCs, HVC~291+26+195 and HVC~297+09+253, were observed 
and studied in high-resolution by BBKW06. Both clouds are associated with the LA. 
BBKW06 analyzed the data  
obtained from the Parkes \HI\ survey of the Magellanic System \citep{Bruns05} 
and interferometer data observed with the 750D configuration of the ATCA. 
By comparing with our 
integrated \HI\ column density map of the ATCA data alone (not shown here), we 
find similar resolved structures as seen in Figure~4 of BBKW06. 
They identified seventeen clumps, which is more than detected in our study,  
although the method of identification 
is different. They did not analyze the combined Parkes and ATCA data. 

Another high-resolution study of a compact HVC (HVC~289+33+251) 
has been carried out by \citet{Bruns04}. This cloud is also 
in the vicinity of LA but a distance of 150~kpc was assumed, which is further 
than the typical distance of the Magellanic Clouds (50~kpc). This cloud 
is significantly brighter than our observed targets and has the typical 
head-tail morphology. It is relatively compact as well. A direct quantitative 
comparison cannot be achieved due to different resolutions 
of the two studies.  

\section{DISCUSSION\label{diss}}

\subsection{HVC Phase Diagram\label{phase}}

HVCs may exist as a stable two-phase medium if the halo pressure lies 
within a certain range, $P_{\rm min} < P < P_{\rm max}$, 
where $P_{\rm min}$ and $P_{\rm max}$ 
are a function of height, $z$, above the Galactic plane \citep{Wolfire95}. 
The phase diagram of thermal pressure, $P$, vs. hydrogen density, $n$, can be used 
to examine the stability of HVCs. To calculate the 
thermal pressure of the resolved clumps, we apply the ideal gas law, $P$/k = $nT_{\rm k}$,  
where k is the Boltzmann constant, $n$ is the hydrogen density and $T_{\rm k}$ is  
the kinetic gas temperature. The upper limit of the 
kinetic gas temperature can be derived from the velocity linewidth assuming 
the observed linewidth is dominated by Doppler broadening, 
$T_{\rm k} = m_{\rm H}\Delta v^{2}$ / (8~k~$\ln$2) = 21.8 ($\Delta v/$\kms)$^{2}$, 
where $m_{\rm H}=1.674\times10^{-27}$ kg. 
Assuming spherical symmetry for the clump, $n = N_{\rm HI}$ / ($d$ tan$\theta$), where $\theta$ 
is the angular diameter of the clump and $d$ is the distance to the cloud. 
The hydrogen density is calculated by assuming distances to the cloud of 25~kpc and 50~kpc. 
The former distance is close to the kinematic distance derived in \citet{NM08}. 
The latter is close to a the measured distance of the LMC \citep{P13}. 
Calculated values are presented in Table~\ref{clumps_tab}. 

\begin{figure}
\includegraphics[scale=0.3]{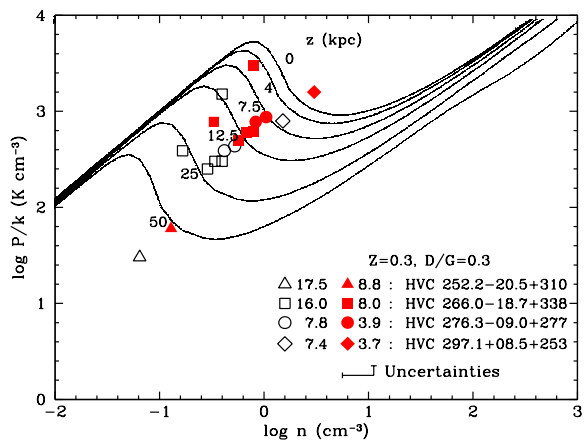}
\caption{Thermal pressure, $P$/k, vs. hydrogen density, $n$, 
for various heights, $z$, above the Galactic plane in kpc. The model of $Z = D/G =0.3$ 
is taken from \citet{Wolfire95}. Filled and non-filled symbols represent 
the calculated data points by assuming a distance of 25 and 50~kpc, 
respectively. The value next to the symbols in the legend are 
the calculated heights $z$, in kpc.}
\label{phase_dia}
\end{figure}

Figure~\ref{phase_dia} shows the phase diagram at different heights $z$ above the Galactic plane 
overplotted with our derived $P$/k and $n$ for the resolved clumps. 
Filled and non-filled symbols correspond to the calculated 
data points by assuming a distance of 25 and 50~kpc, respectively. 
The inferred $z$ with the assumed distance for each HVC clump is listed next to each symbol.
We note that the phase diagram 
is sensitive to the metallicity ($Z$) and dust-to-gas ratio ($D/G$). 
We adopt the model of $Z = D/G = 0.3$, which is appropriate for stripped LMC 
gas \citep{Wolfire95}.  

To estimate the effects of uncertainties in our derived $P/k$ and $n$, we consider various 
scenarios. Assuming the observed clump is not of spherical shape but an ellipsoid, 
viewing along the major axis gives a larger angular diameter. 
Twice the size in angular diameter will result in a 
decrease in hydrogen gas density by a factor of two. 
The derived $T_{\rm k}$ is an upper limit given by the velocity linewidth. 
A decrease of $T_{\rm k}$ by 100~K will result in 
+0.1 change in $\log P$/k. Ignoring the two data points near the 50~kpc curve and taking 
into account the uncertainties, we find that majority of the data points fall onto 
their corresponding $z$ for the phase diagram. Also, the data points lie in the instability 
valley of the two-phase medium.

\subsection{Halo Environment as a Function of Galactic Latitude\label{haloenv}}

\begin{figure}
\includegraphics[scale=0.45]{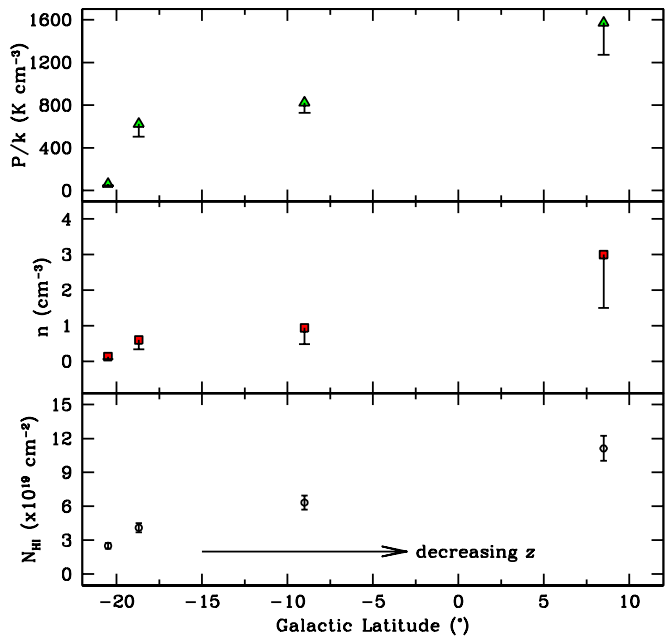}
\caption{Thermal pressure, $P$/k, hydrogen density, $n$, and 
\HI\ column density, $N_{\rm HI}$, as a function of Galactic latitude (as a proxy of $z$). 
Each data point represents the average physical parameters of 
all resolved clumps for each cloud. A distance of 25~kpc is assumed here. 
Uncertainties of $P$/k and $n$ are discussed in 
\S7.1. A systematic uncertainties of 10$\%$ is 
assumed for $N_{\rm HI}$. Significantly deviating values, such as 
the ones for clump D and E of HVC~266.0-18.7+338, are excluded from the calculation.   
Physical parameters are listed in Table~\ref{clumps_tab}.}
\label{gal_par}
\end{figure}

If we assume that the clumps are in hydrostatic equalibrium, 
the thermal pressure of the clump is equal to the external halo thermal pressure. 
In Figure~\ref{gal_par}, we show the average physical parameters of the resolved clumps 
for each cloud as a function of Galactic latitude. In this case, 
Galactic latitude is a proxy for $z$. The height $z$ decreases from negative to positive 
Galactic latitude. Interestingly, 
we find a trend of increasing halo thermal pressure, hydrogen density and 
\HI\ column density with increasing Galactic latitude. This suggests 
that the clouds reside in a denser halo environment at more positive Galactic latitude.
While it has been suggested that the LA is closer to the Galactic centre than the LMC, 
\citep{NM08}, this is the first evidence showing a possible distance gradient in the LA region 
using compact HVCs that span over $\sim80\degr$ in Galactic longitude.
The leading part of the LA (LA II and LA III) is most likely closer to our Galaxy 
than the LA I, which has a kinematic distance of $\sim$21~kpc \citep{NM08}. 
The study of \citet{Venzmer12} also showed a 
similar distance gradient using a different approach. 
Their investigation was based on the velocity structure of three subpopulations in LA~I.
Future simulations would be useful to assess the likelihood of such a scenario. 
We note that 
$P$/k and $n$ are sensitive to various scenarios as discussed in \S7.1. 
The positive Galactic latitude cloud (HVC~297.14+08.5+253) will have a lower $n$ if 
it is not spherical but twice the size. The trend of $n$ vs. Galactic latitude 
becomes less significant in this case.

In FSM13, the formation of the LA IV remains a mystery. 
It has a very different morphology than its counterpart (LA~I--III). 
This suggests that LA IV might be formed via a different mechanism 
or has a different origin. The recent discoveries of ultra-faint dwarf galaxies in the 
vicinity of the Magellanic System \citep{Koposov15,Bechtol15,D15}
has renewed theoretical and observational interest (see \citealp{C15,Westmeier15}). 
The model of \citet{C15} confirms that the locations of these ultra-faint dwarf galaxies are associated 
with the MCs previously as part of a loose group and shows 
how they are processed by the Galactic halo upon accretion. 
This suggests that the clumpy LA~IV might be the debris from such an accretion event, which 
might explain why the three HVCs (part of the LA~IV) are further away than 
the two HVCs near LA I and II. This supports the radial distance estimate of 74~kpc for LA~IV in \citet{Venzmer12}.

\subsection{Star Formation in the Leading Arm Region\label{starform}}

Theoretical models and observational evidence show that both the MS and LA were 
formed via ram-pressure stripping and tidal interaction between the LMC and SMC  
(see e.g. \citealp{DB11,For13}). Both of these mechanisms are known to 
trigger star formation in galaxies (e.g., galaxy pair NGC~1512/1510, 
\citealp{Koribalski09} 
and NGC~4522 in the Virgo cluster, \citealp{Kenney04}). 
Star formation in the MS and LA has been postulated 
but there was no success in the hunt for stars being formed in situ in early years 
(e.g., \citealp{RC82,GR98}). \citet{DM09} conducted a 
search for star formation in 
the MS regions using cold atomic gas as an indicator. 
Two absorption components were identified toward 
a background radio source, J0119$-$6809. The corresponding \HI\ column density is 
$\sim2\times10^{20}$~cm$^{-1}$. Follow up observation shows 
no detection of CO($J=1\rightarrow0$) molecular gas associated 
with the cool gas implying that star formation does not occur at that location within the MS. 

Neutral hydrogen gas, dust and molecular hydrogen gas (H$_{\rm 2}$) act as a reservoir for 
fuelling star formation activities. This is particularly prominent 
in high gas density regions. Stars are formed when the dense gas clouds 
reach the Jeans instability and collapse. 
Measurements of dust and H$_{\rm 2}$ are fairly limited in the LA region. 
An attempt for measuring dust, H$_{\rm 2}$ and metallicity 
in the LA has been made by \cite{Sembach01} on 
a compact HVC (HVC~287.5+22.5+240). They 
find that the metallicity of this HVC is similar to the SMC, and 
the detection of 
H$_{\rm 2}$ suggests that either the H$_{\rm 2}$ formed in situ or within the 
SMC and survived tidal stripping. They prefer the latter scenario given that the formation 
timescale of H$_{\rm 2}$ is long ($\sim10^{8}$ yr). 
If the LA is stripped from the SMC as suggested by simulations  
(e.g., \citealp{DB11}), we would expect the LA region has a similar 
H$_{\rm 2}$ and metallicity content as the SMC, and star formation potentially could occur. 
In fact, a recent attempt to search for 
stellar components has been carried out at the optical wavelength by \citet{CD14} 
in the LA region. Five young stars has been successfully identified 
and they believed to have  
formed in the LA for the first time, 
using kinematics information, stellar parameters and distance moduli. 
We examine the location of these 5 young stars and nearby \HI\ gas content. 
Four of them are in close proximity of dense clouds with $N_{\rm HI}$ 
in the range of $1.5\times10^{19}$ to $1.8\times10^{20}$~cm$^{-2}$. 
Clouds with these \HI\ column densities exist everywhere in the LA region and yet the 
search only resulted in five young stars that were born in situ. Where 
are the missing stars? 

The result from \S\ref{phase} suggests the possibility of 
different halo environments in the LA region, which might explain the lack of 
star formation if a special condition is required to form stars. 
However, without further information on dust and metallicity 
properties in the entire LA region, it is very hard to assess the conditions 
that are needed for star formation to occur. 
The optical search for young stars is currently 
limited to specific regions. 
It would be interesting if future optical studies included candidates located 
in the vicinity of HVC~287.5+22.5+240 to verify if star formation is plausible 
with the given conditions. 

Metallicity and dust might not be the only factors for the star formation in 
the region. Observational evidence of interaction between the 
LA~I and Galactic disc gas suggests  
the gas is being compressed by the Galactic halo gas at low $z$ and star formation could also be 
triggered.

\section[]{SUMMARY\label{sum}}

We have studied five HVCs in the vicinity of the Magellanic Leading Arm. 
The targets were selected from the FSM13 catalog, and high-resolution 
observations 
were carried out at the ATCA. We analyzed the combined 
single-dish GASS and interferometric ATCA data. Clumps were identified 
and physical parameters were derived for both clumps and 
diffuse structure of the HVCs. 
Most of the clumps have a cold component ($\Delta v < 10$~\kms). 
The unresolved clumps generally consist of a warm component only. 

Three of the clouds are part of the LA IV, which lies south of 
the Galactic plane, 
and have a head-tail like morphology. 
The other two are located 
north the Galactic plane and in close proximity to LA~I and LA~II. 
HVC~266.0-18.7+338 is the only cloud that shows a clear velocity gradient. 
It also consists of many clumps. 
The $V_{\rm LSR}$ of cold clumps is generally larger than the diffuse (warm) component. 

In the case of HVC~276.3-09.0+277, its head is compressed and 
it does not show any velocity gradient. The velocity of the cold component decreases 
gradually and is larger than that of warm component. Two clumps are located in the two main 
cores of the cloud. The third clump is relatively small and unresolved. It is 
located slightly offset from one of the clumps. 

HVC~297.1+08.5+253 has been studied by BBKW06. They analyzed the 
single-dish and interferometer data separately. Their analysis focuses on the 
core of the head-tail structure and many clumps have been identified. 
We, on the other hand, carried out the analysis by using the combined image. 
This allows us to probe large and small scale structures simultaneously. 
In our analysis, the $N_{\rm HI}$ increases toward the head of the cloud 
and decreases afterward. The cold component shows the same $N_{\rm HI}$ pattern except 
that it peaks at a slightly different position from the warm component. There is a small 
velocity gradient at the tail of the cloud. Otherwise, velocity is constant for both 
cold and warm components. 

The analysis of the combined image of HVC~310.3+08.1+167 
was carried out on a smoothed cube due to it being very diffuse and low 
in surface brightness. It consists of many clumps, and all of them are 
unresolved with cold components. Overall, the velocity linewidth of both the cold and 
warm component follows the same trend as the $N_{\rm HI}$.

We discussed the HVC phase diagram (log $P/k$ vs log $n$) using the model of \citet{Wolfire95}. 
The model has metallicity and gas-to-dust ratio of 0.3, which is based on the 
assumption that the gas was stripped from the LMC. 
All the clouds are in the instability valley where they can maintain their two-phase structures. 
Interestingly, there is an offset between the data points and 
their corresponding height above the Galactic plane. 
A lower metallicity model is a better fit for the majority of the data points, 
which suggests that the gas was likely stripped from the SMC. This is consistent with 
various simulations (e.g. \citealp{DB11}). 
We find a gradient in thermal halo pressure, hydrogen density and \HI\ column density 
as a function of Galactic latitude. This is the first possible observational evidence of 
an increasing Galactocentric distance from the trailing end to the leading part of the LA, although further observations may be needed to confirm.
A different halo environment might explain 
the low star formation rate in the LA region. Special conditions such as 
additional dust and atomic hydrogen gas are needed in order 
to trigger star formation. Future studies of the 
atomic hydrogen gas content in the region and the search for more stars 
will be important to understand the star formation history in the region of the LA.

\section*{Acknowledgments}

B.-Q. F. was the recipient of a John Stocker Postdoctoral Fellowship 
from the Science and Industry Research Fund. 
N. Mc-G acknowledges CSIRO Astronomy \& Space Science where this work was commenced.
This publication made use of data products from the Parkes and ATCA radio telescopes. 
The Australia Telescope Compact Array/Parkes radio telescope is part of the 
Australia Telescope National Facility, which is funded by the 
Commonwealth of Australia for operation as a National Facility managed by CSIRO. 

\bibliographystyle{mnras}
\bibliography{ref}

\end{document}